\documentclass[12pt]{article}

\usepackage{hyperref}

\usepackage{epsf}
\usepackage{amsmath,amssymb}
\usepackage{latexsym}
\usepackage{graphicx,color}
\usepackage{wrapfig}
\usepackage{latexsym}
\usepackage{graphics}

\usepackage{color}
\usepackage{delarray,color}
\usepackage{fancybox}  

\newcommand {\beq}{\begin{align}}
\newcommand {\eeq}{\end{align}}

\newcommand{\be}{\begin{equation}}
\newcommand{\ba}{\begin{align}}
\newcommand{\ea}{\end{align}}
\newcommand{\ee}{\end{equation}}

\newcommand{\beqa}{\begin{align}}
\newcommand{\eeqa}{\end{align}}
\newcommand{\CR}{\nonumber \\}

\newcommand{\unit}{\hbox to 3.8pt{\hskip1.3pt \vrule height 7.4pt
    width .4pt \hskip.7pt \vrule height 7.85pt width .4pt \kern-2.4pt
    \hrulefill \kern-3pt \raise 3.7pt\hbox{\char'40}}}

\def\matt[#1,#2,#3,#4]{\left(%
\begin{array}{cc} #1 & #2 \\ #3 & #4 \end{array} \right)}

\usepackage{tabularx}

\catcode`\@=11
\@addtoreset{equation}{section}

\catcode`@=12
\relax

\begin{document}

\begin{titlepage}

\setcounter{page}{0}

\renewcommand{\thefootnote}{\fnsymbol{footnote}}

\begin{flushright}
YITP-18-95 \\
\end{flushright}

\vskip 1.35cm

\begin{center}
{\Large \bf 
Supersymmetry Breaking Phase \\ \vspace{2mm}
in Three Dimensional Large $N$ Gauge Theories
}

\vskip 1.2cm 

{\normalsize
Kazuma Shimizu\footnote{kazuma.shimizu(at)yukawa.kyoto-u.ac.jp} and Seiji Terashima\footnote{terasima(at)yukawa.kyoto-u.ac.jp}
}

\vskip 0.8cm

{ \it
Yukawa Institute for Theoretical Physics, Kyoto University, Kyoto 606-8502, Japan
}

\end{center}

\vspace{12mm}

\centerline{{\bf Abstract}}

Three dimensional supersymmetric gauge theories
are often in a gapped
phase, in which SUSY is 
spontaneously broken, if all the matter fields are massive 
and decoupled in the low energy.
We study this phase in the large $N$ limit
using the localization technique for the theory on 
the ellipsoid, which interpolates the round three sphere and
the flat space compactified on $S^1$.
We find a large $N$ saddle point solution
for the gauge theory with some massive matter fields.
This solution gives 
a vanishing (generalized) Polyakov loop in the flat space limit,
thus, it corresponds to the confining phase
at the leading order 
in the $1/N$ expansion.

\end{titlepage}
\newpage

\tableofcontents
\vskip 1.2cm 

\section{Introduction}

For a supersymmetric (SUSY) gauge theory on a compact space,
we can apply the localization technique to
compute some SUSY invariant quantities exactly \cite{Pestun, Witten, Nekrasov}.
In particular, 
for ${\cal N}=2$ three dimensional SUSY gauge theories on $S^3$,
the partition function and the SUSY Wilson loop can be
expressed by the matrix model type integration, instead of the path-integral \cite{KWY1, Jafferis, Hama}.
There are many interesting phenomena in three dimensional SUSY gauge theories,
which have been studied by the localization technique. 
On the other hand, 
in the three dimensional SUSY gauge theory SUSY, is often broken spontaneously. 
Indeed, 
the ${\cal N}=2$ SUSY Chern-Simons theory
is believed to be in the mass gapped phase, in which 
SUSY is 
spontaneously broken, for $N > k$ where 
$k$ is the Chern-Simons level and
the gauge group is SU($N)$  \cite{IS, Witten1, Ohta}.
This phase will be realized for varieties of the 
three dimensional SUSY gauge theory if all of the matter fields are massive
and can be regarded to be decoupled.
On the flat space, such decoupling of matter fields occurs 
at the origin of the Coulomb branch 
which may be regarded as the metastable vacuum.

In the large $N$ limit with finite $k$,
the topological degrees of freedom in the low energy 
may not be relevant in the leading order in the $1/N$ expansion
and the phase will be considered as the SUSY breaking phase
in the large $N$ limit.
It may be interesting to study this phase through the exact results
obtained by the localization technique
because such a phase can appear in many interesting models.
Indeed, such a spontaneously broken SUSY phase has been argued 
to appear in the large $N$ 
mass deformed ABJM theory on $S^3$,
which will have a gravity dual with an asymptotic AdS$_4$ geometry,
for an enough large mass parameter \cite{NST2,HNST}.

In this paper, we study this phase in the large $N$ limit
with finite $k$,
using the localization technique for the theory on 
the ellipsoid, which interpolates the sphere and
the flat space compactified on $S^1$.
Because of the large $N$ limit, 
we can evaluate the matrix model integral 
by the saddle point approximation.
We find a large $N$ saddle point solution
for the gauge theory (with massive matter fields which transform as 
the adjoint representation of the gauge group).
The solution gives the vanishing free energy in the leading 
order in $1/N$ expansion except for the contributions from 
the decoupling matter fields.
This indicates that the ${\cal O}(N^2)$ gluons are 
confined.
We also see that the solution is consistent 
with the exact results of the
low energy ${\cal N}=2$ SUSY Chern-Simons theory
which is believed to be in a SUSY breaking phase.

The SUSY Wilson loop on the ellipsoid 
for this solution is also computed 
and shown to vanish in the leading 
order in $1/N$ expansion.
In the flat limit of the ellipsoid, 
the Wilson loop can be regarded as
(SUSY generalized) Polyakov loop.
Thus, the solution corresponds to the confining phase
(i.e. the center symmetry preserving phase)
in the large $N$ limit.
This result is somewhat surprising
because the localization technique reduces
the path integral variables to the integrations over 
the constant scalars where the gauge fields are 
fixed to zero
although
${\mathbb Z}_N$ symmetric configurations for the non-vanishing gauge fields
give the vanishing Polyakov loop.
In our solution, the nonzero imaginary 
scalars values are similar to the ${\mathbb Z}_N$ symmetric configurations for the gauge fields.
Furthermore, the scalars and gauge fields
are combined like complex variables in the SUSY Wilson loop,
then the vanishing (generalized) Polyakov loop is realized.

The theory which we will expect to have such a SUSY breaking phase is the mass deformed ABJM theory
with a sufficiently large mass.
In \cite{NST1}, \cite{NST2} and \cite{HNST}, it has been argued that
there is a critical mass in this theory and 
if the mass is larger than this critical value, the SUSY breaking occurs.
In this phase, 
it is possible that 
the large $N$ solution discussed in this paper is valid.
One thing we need to consider for the mass deformed ABJM theory
is the critical value of the mass for the theory on the ellipsoid.
In order to do this we study the large $N$ solution for a small mass
on the ellipsoid.

The rest of paper is as follows: In section 2, we introduce the ellipsoid $S^{3}_{b}$ and see that the manifold is regarded as the $S^{1}\times \mathbb{R}^2$ in a large $b$ limit. 
Then we review some ingredients of the localization technique and apply it to the supersymmetric gauge theory on $S^{3}_{b}$. In section 3, we investigate large $N$ solutions of several kinds of gauge theories and show that there exists the solution corresponding to the SUSY breaking phase. In section 4, we solve the saddle point equation of the mass deformed ABJM theory on $S^{3}_{b}$ and find 
the critical mass discussed in \cite{NST1, NST2} for $S^{3}_{b}$. In section 5 we summarize this paper and discuss some problems. In appendix A, we discuss which solution is dominant in the large $N$ limit. We introduce the theory whose saddle point equation has at least two types of the solutions. We investigate when the solution in the SUSY breaking phase tends to become the dominant one.

\section{Three dimensional ellipsoid $S^3_{b}$}

The metric of the 3d ellipsoid $S^3_{b}$ is 
\begin{align}
 ds^2 = l^2 \left(
\frac{1}{b^2} \cos^2 \theta d \varphi^2 +b^2 \sin^2 \theta d \chi^2+f^2 d \theta^2,
\right)
\end{align}
where $0 \leq \theta < \pi/2$, $0 \leq \varphi < 2 \pi $, $0 \leq \chi < 2 \pi $ and 
\begin{align}
 f^2=\frac{\sin^2 \theta}{b^2} +b^2 \cos^2 \theta.
\end{align}
Later, we will consider the flat limit in which we take
\begin{align}
 b \rightarrow \infty, \,\,\,
l \rightarrow \infty, \,\,\,\,
l_1 \equiv \frac{l}{b} = {\rm finite},
\end{align}
and consider only the region $\theta \ll 1$ with
\begin{align}
 r \equiv b^2 \theta ={\rm finite}.
\end{align}
In this limit, 
we have
\begin{align}
 ds^2 \rightarrow (l_1)^2 \left( d \varphi^2 + r^2 d \chi^2 + dr^2 \right),
\end{align}
which is $S^1_{l_1} \times {\mathbb R}^2$, near $\theta =0$.
The theory in the gapped phase with the gap $E$ is isolated 
to be in this geometry near $\theta =0$ 
if $ l \gg 1/E$ except the remaining topological degrees of freedom.

Below, we will set $l=1$ for notational simplicity.
We can recover $l$ dependence by the dimensional analysis.

Let us consider the SUSY gauge theory on it.
For this, we need the non-zero back ground gauge field
\begin{align}
 V=-\frac{1}{2} \left( 1- \frac{1}{ bf} \right) d \varphi
+\frac{1}{2} \left( 1- \frac{b}{ f} \right) d \chi,
\end{align}
which couples to the R-symmetry current.
Then, a natural choice for the dreibein is 
\begin{align}
e^1=\frac{1}{b} \cos \theta d \varphi, 
\,\,\,
e^2=b \sin \theta d \chi, 
\,\,\,
e^3= f d \theta,
\end{align}
and the ``Killing spinors'' of the SUSY generators can be taken as
\begin{align}
 \epsilon = \frac{1}{\sqrt{2}} 
\left(
\begin{array}{c} -e^{\frac{i}{2}(\chi-\varphi+\theta)}  \\ 
e^{\frac{i}{2}(\chi-\varphi-\theta)} 
  \end{array}
\right), \,\,\,\,\,
 \bar{\epsilon} = \frac{1}{\sqrt{2}} 
\left(
\begin{array}{c} e^{\frac{i}{2}(-\chi+\varphi+\theta)}  \\ 
e^{\frac{i}{2}(-\chi+\varphi-\theta)} 
  \end{array}
\right),
\end{align}
where the R-charges of $\epsilon$ and $\bar{\epsilon}$
are $1$ and $-1$, respectively.
They satisfy $D_m \epsilon=\frac{i}{2 f} \gamma_m \epsilon $
and $D_m \bar{\epsilon}=\frac{i}{2 f} \gamma_m \bar{\epsilon} $.
In the flat limit near $\theta=0$, 
the background becomes 
\begin{align}
 V \rightarrow -\frac{1}{2} d \varphi,
\label{bV2}
\end{align}
which implies $dV=0$, and
\begin{align}
 \epsilon \rightarrow  \frac{1}{\sqrt{2}} 
e^{\frac{i}{2}(\chi-\varphi)} \left(
\begin{array}{c} -1 \\ 1
  \end{array}
\right), \,\,\,\,\,
 \bar{\epsilon} = \frac{1}{\sqrt{2}} 
e^{\frac{i}{2}(-\chi+\varphi)} \left(
\begin{array}{c} 1  \\ 1
  \end{array}
\right),
\end{align}
which are indeed constant spinors in 
$S^1 \times {\mathbb R}^2$ with the 
factor $e^{\pm i\varphi/2}$ coming from the background (\ref{bV2}),
where $\chi$ dependence comes from the local Lorentz transformation.
They satisfy $D_m \epsilon=0$ and $D_m \bar{\epsilon}=0$.

\subsection{Localization}

The partition function and some Wilson loops 
of any 3d ${\cal N}=2$ SUSY gauge theory 
on the ellipsoid $S^3_b$
can be computed exactly by the localization technique \cite{Hosomichi}.
First, we will review shortly the results in \cite{Hosomichi,Kapustin,3d}.
The saddle points are 
$F_{mn}=D_m \sigma=D=\phi=\bar{\phi}=F=\bar{F}=0$,
which mean that 
only scalars $\sigma$ in the vector multiplets
can have non-zero values.
For the saddle points, they should be 
covariantly constant and we will denote them as
\begin{align}
 \sigma|_{\rm saddle}=a,
\end{align}
which is in a Cartan sub-algebra by the gauge transformation.
Then, for the saddle points 
the Chern-Simon terms and the FI term are evaluated 
as
\begin{align}
 e^{-S_{\text{Chern-Simons}}}=e^{i \pi k {\rm Tr} (a^2)}, \,\,\,
e^{-S_{\text{FI}}}=e^{4 \pi i \zeta \text{Tr}(a)}.
\end{align}
The Yang-Mills terms for the vector multiplets 
and the kinetic terms and the superpotential terms for the chiral
multiplets can not contribute to the partition function at least classical level.
The 1-loop factor for the vector multiplet with the integration measure 
for $a$ is given by
\begin{align}
 \frac{1}{|\cal W|} \prod_{i=1}^r d a_i \prod_{\alpha \in \Delta_+}
4 \sinh (\pi b \, a \cdot\alpha) \sinh(\pi \frac{1}{b} a \cdot \alpha),
\end{align}
where $G$ is the gauge (simple) group, $|\cal W|$ is the order of the Weyl group
and $\Delta_+$ is the set of the positive roots.
For the chiral multiplet whose bottom component has R-charge $r$, the 1-loop factor is 
\begin{align}
 \prod_{w \in R } s_b (\frac{iQ}{2} (1-r) - a \cdot w),
\end{align}
where $w$ runs the weights in the representation of $G$ for the chiral multiplet,
\begin{align}
 Q=b+\frac{1}{b},
\end{align}
and
\begin{align}
 s_b(z) \equiv \prod_{m,n=0}^\infty 
\frac{m b +n b^{-1}+\frac{Q}{2}-iz}{m b +n b^{-1}+\frac{Q}{2}+iz},
\end{align}
is the double sine function. We introduce some properties of the function studied in \cite{BT,Hatsuda}. $s_{b}(x)$ satisfies 
\begin{align}
 s_b(z)= s_{b^{-1}} (z), \,\,\,  
 s_b(z) s_b(-z)=1,
\end{align}
and the expansion around ${\rm Re} (z) =\infty$:
\begin{align}
i \log s_b(z) = - \frac{ \pi z^2}{2} - \frac{\pi}{24} (b^2+b^{-2})
+\sum_{l=1}^{\infty} \frac{(-1)^{l-1}}{l} 
\left(
\frac{e^{- 2\pi l bz}}{2 \sin(\pi l b^2)} +
\frac{e^{- 2\pi l z/b}}{2 \sin(\pi l b^{-2})} 
\right),
\label{ex1}
\end{align}
and around ${\rm Re} (z) =-\infty$:
\begin{align}
i \log s_b(z) =  \frac{ \pi z^2}{2} + \frac{\pi}{24} (b^2+b^{-2})
+\sum_{l=1}^{\infty} \frac{(-1)^{l}}{l} 
\left(
\frac{e^{ 2\pi l bz}}{2 \sin(\pi l b^2)} +
\frac{e^{ 2\pi l z/b}}{2 \sin(\pi l b^{-2})} 
\right).
\label{ex2}
\end{align}
For the vector like matters, the 1-loop factor becomes
\begin{align}
\left( 
\prod_{w \in R } s_b (\frac{iQ}{2} (1-r) - a \cdot w)
\right)
\left( 
\prod_{w \in R } s_b (\frac{iQ}{2} (1-r) + a \cdot w)
\right)
=\prod_{w \in R } D_{-iQ(1-r)/2} (a \cdot w),
\end{align}
where 
\begin{align}
 D_\alpha(x) \equiv \frac{s_b(x-\alpha)}{s_b(x+\alpha)}.
\end{align}
This function satisfies 
\begin{align}
 D_\alpha(x)= D_\alpha(-x).
\end{align}
We also define 
\begin{align}
D_b(x) \equiv  D_{-iQ/4}(x),
\end{align}
which satisfies $D_b(x)|_{b=1}=\frac{1}{2 \cosh (\pi x)}$. 
When $|\text{Im}(x)|<\frac{|\text{Re}Q|}{2}$, 
$\log D_{b}(x)$ has the following integral form:
\begin{align}
\label{integral}
\log D_{b}(x)=\int_{\mathbb{R}+i0}\frac{dt}{2t}\frac{\sinh \left(\frac{Qt}{2}\right)\cos(2xt)}{\sinh(bt)\sinh(b^{-1}t)}.
\end{align}

Another useful formula we will use later is 
the following expansion for  $x$ with a large positive real part:
\begin{align}
 \log D_b (x) = - \frac{\pi Q}{2} x
+\sum_{n=1}^\infty 
\left(
\frac{e^{-2 \pi n b x}}{2 n \cos ( \frac{\pi n b Q}{2})}
+\frac{e^{-2 \pi n b^{-1} x}}{2 n \cos ( \frac{\pi n b^{-1} Q}{2})}
\right).
\end{align}

The SUSY Wilson loop on $S^1$ at $\theta=0$ 
\begin{align}
W_R(\theta=0) \equiv {\rm Tr}_R {\rm P} \exp 
\oint_{\theta=0} (i A+\sigma dl),  
\label{SW}
\end{align}
is also evaluated by the localization technique. At the saddle points,
the Wilson loop becomes
\begin{align}
W_R(\theta=0)|_{\rm saddle} ={\rm Tr}_{R} (e^{2 \pi \frac{a}{b}}).  
\end{align}

This will be considered as a generalized Polyakov loop
in the $b \rightarrow \infty$.
This wraps around the $S^1$ in $S^1 \times {\mathbb R}^2$.
There are some differences between the usual Polyakov loop
and the one considered in the paper.
The theory is on $S^1 \times {\mathbb R}^2$ instead of $S^1 \times {\mathbb R}^3$ and
there is the non-zero background gauge field corresponding
to the R-charge.
Furthermore, the SUSY Wilson loop (\ref{SW}) includes 
the scalar $\sigma$.
Although these differences exist, 
the SUSY Wilson loop (\ref{SW}) can be regarded as an external particle of the representation $R$
on $S^1$. 
Furthermore, there is the center symmetry ${\mathbb Z}_N$ in the flat space limit and
the usual Polyakov loop is the order parameter of this symmetry.
We expect that the SUSY Wilson loop (\ref{SW}) also is the order parameter of the
center symmetry because 
the center symmetry does not act the scalar in it and
the SUSY Wilson loop is expected to be transformed under the symmetry as the non-SUSY Wilson-loop. 
Thus, if $ \langle W_{R} \rangle \neq 0$ the center symmetry is spontaneously broken
although the SUSY and no SUSY Wilson loops will take different values.
Note that only the ${\mathbb R}^2$ is the non-compact space, 
thus there are no spontaneous breaking of symmetries if $N$ is finite.
However, in this paper, we consider the large $N$ limit 
of the theory and thus symmetries can be broken spontaneously.
Therefore, we will call 
the phase with $ \langle W_{R} \rangle =0$  
``confinement phase'' although 
it is only meaningful for the large $N$ limit.

\section{Large $N$ analysis of ${\cal N}=2$ gauge theories on $S^3_b$}

In this section, using the localization results,
\begin{align}
 Z=\int \prod_{i=1}^r d a_i \, e^{-S(a)},
\label{mma}
\end{align}
we will compute the partition function and 
the Wilson loop in the large $N$ limit
in which we keep other parameters 
(the Chern-Simons level $k$, FI parameter or mass, $b$ and 
the length  scale $l$) 
finite.
In the limit, 
the integrations over the localization saddle points $a_i$
may be dominated by the large $N$ saddle points which are
given by
\begin{align}
 \frac{\partial S(a)}{\partial a_i}\big|_{\rm leading}= 0,
\label{se}
\end{align} 
where $|_{\rm leading}$ means taking  
the large $N$ leading order part.

Below, we will take the gauge group G=U($N)$
and denote $a \cdot \alpha_{ij}=a_i-a_j$ with
$i,j=1, \ldots, N$ for the adjoint representation\footnote{
In the large $N$ analysis in this paper can be 
trivially extended to SU($N)$ case by 
imposing the condition $\sum_{i=1}^N a_i=0$
and the results will not change.}.

\subsection{Large $N$ solution for confinement phase}

Here we investigate the existence of the solution corresponding to the confinement phase in various theories on ellipsoid $S_{b}^{3}$.\footnote{ 
This type of the solution exists even in the theory whose matrix model does not converge.
In that case, it is not guaranteed that the solution is meaningful. 
}
Depending on the theory with the parameters, there is another solution
which approximates the matrix integral in the large $N$ limit instead of the solution for the confinement phase.  
In this section, we will not discuss which solution indeed approximates the matrix integral.
However, for the mass deformed ABJM, which is discussed in the next section, 
we argue that the solution for the confinement phase can be appropriate for large enough mass.
For the theories with massive adjoint matter fields
with mass $m$, 
we will discuss this problem in the appendix
and see that the large $N$ solution is expected to be 
valid if $km/N$ is finite and large enough.

\subsubsection{Pure ${\cal N}=2$ SUSY Chern-Simons Yang-Mills theory}

First, we consider the theory without chiral multiplets\footnote{
In this case, although the integrations over $a_i$ diverge,
even for non-zero $k$ without a precise regularization.
We will analyze them assuming the regularization is done.
In fact we can introduce the imaginary part of the Chern-Simons-level for the matrix model to converges and finally take it to zero. The existence of this imaginary part of the Chern-Simons-level does not change the solution of the saddle point equation because the solution we will consider does not depend on the Chern-Simons-level. In the recent works \cite{CKW1, CKW2}, it was shown that we can take the contour so 
that the integrations of the matrix models converge.}. The matrix model action $S$ in (\ref{mma}) becomes
\begin{align}
S(a) =& i \pi k \sum_{i=1}^N a_i^2 - 2 \pi i \zeta  \sum_{i=1}^N a_i \CR
&-\frac{1}{2} \sum_{i,j=1, \,\, i>j} 
\left( \log 4 \sinh (\pi b (a_i-a_j)+\log 4 \sinh (\pi b^{-1} (a_i-a_j)
\right)+N \log N,
\label{sa1}
\end{align}
up to a constant.\footnote{
We have approximated $\log N! \approx N \log N$.}
In particular, for $k=0$, the matrix integral is not convergent and 
the theory is the
pure ${\cal N}=2$ SUSY Yang-Mills theory which has the runaway type effective potential \cite{AHW}.
The saddle point equation is written down as
\begin{align}
\label{saddis}
0=\frac{\partial S(a)}{\partial a_{j}}=2i\pi k a_{j}-2i\pi \zeta-b\pi\sum_{k\neq j}\coth\pi b(a_{j}-a_{k})-b^{-1}\pi\sum_{k\neq j}\coth\pi b^{-1}(a_{j}-a_{k}).
\end{align}
Let us take the special class of the ellipsoids which have
\begin{align}
 b =\sqrt{\frac{p}{q}}, \,\,\,\,
\end{align}
where $p,q \in {\mathbb Z}$.
For these special values, we will find 
large $N$ solutions corresponding to
the confinement phase\footnote{
It is interesting to extend the solution without this 
conditions on $b$. It is also interesting to find the reason why this condition should be imposed.
By now, we do not have any answers.
}.

Extending the solution given in \cite{NST2},
the solution is given by
\begin{align}
 a_j=i \left( \frac{j}{N} -c
\right) \sqrt{p q}M,
\label{s1}
\end{align}
where $M= {\mathbb Z}_{>0}$ and $c=\frac{N+1}{2 N} $.
Note that we took the constant $c$ such that $\sum_{j=1}^N a_j=0$.
For this, the saddle point equation in the large $N$ limit (\ref{se})
becomes
\begin{align}
 0=b \int_0^{1} dy \cot  \pi p  (x-y) 
+b^{-1} \int_0^{1} dy \cot  \pi q  (x-y), 
\label{spe1}
\end{align}
where we assume the continuous limit in which we replace the discrete valuable $j$ and the summation of it from continuous ones $x$ and the integral over $x$ as 
\begin{align}
\frac{j}{N}\rightarrow x\in[0,1],\quad \frac{1}{N}\sum_{j=1}^{N-1}\rightarrow \int_{0}^{1} dx.
\end{align}
The integrals are defined as the principal values because
the zero of the $\sinh$ is not included in (\ref{sa1})\footnote{
Here we assumed $pq$ and $N$ are coprime.
Even if we do not assume this, we expect that
there are appropriate regularizations to avoid the singular point.
}.
Here the Chern-Simons term and the FI term were neglected in the large $N$ limit 
because $k$ and $\zeta$ are finite.
Then, indeed, the integrations over $y$ in (\ref{spe1})
vanish for any $x$ because the integration of the $\cot$ 
over the period vanishes and $p,q$ are integers\footnote{
For this solution, there are infinitely many discrete moduli,
$a_j=i \left( \frac{j}{N} -c+n(j)
\right) \sqrt{p q}M$ where $n(j)$ is an ${\cal O}(N^0)$ integer,
which do not change the values of the free energy and the Wilson loop \cite{NST2}.
The summation over these may not affect the large $N$ 
results 
although these could be important to obtain $Z=0$,
which is expected for the theory in the SUSY breaking,
through the Chern-Simons term.
}.

The v.e.v. of 
the Wilson loop at $\theta=0$ for the fundamental representation 
is computed in the leading order in the large $N$ limit as
\begin{align}
 \langle W \rangle \approx 
e^{- \pi i q }
\sum_{j=1}^N 
\exp \left( 2 \pi i \frac{j}{N} q  \right)=0,
\end{align}
where we neglected the sub-leading term which will be ${\cal O}(N^0)$.
This means that the theory is in the confining phase,
at least in the large $b$ limit which can be taken,
for example, by taking $q=1$ and $p \rightarrow \infty$.

The free energy, which is defined as $F \equiv - \log |Z|$,
is computed for the solution (\ref{s1}) as 
\begin{align}
F = 0 \cdot N^2  + {\cal O}(N).
\end{align}
\begin{align}
\label{lf}
\frac{N^2}{4}\int_{1\geq x>y\geq 0}dxdy\left[\log\sin^2\pi p(x-y)+\log\sin^2\pi q(x-y)\right]=0.
\end{align}
This result also supports that the theory 
is in the confining phase where the ${\cal O} (N^2)$ gluons and other fields
are confined.

Note that at the saddle point,
$\sigma$ is pure imaginary. This would be related 
to the expected value of the usual Polyakov loop
because  the combination of 
the gauge field $i A_\mu$ and $\sigma$ 
appears
in the supersymmetric Wilson loop.

We have considered the large $N$ limit of the
${\cal N}=2$ SUSY Chern-Simons Yang-Mills theory,
however, even for finite $N$ 
the partition function and Wilson loop have been computed.
Under the certain regularization, the partition function for the 
${\cal N}=2$ SUSY SU($N)$ Chern-Simons theory 
on the three-sphere
can be explicitly computed as \cite{KWY1}
\begin{align}
\label{ChernSimons1}
|Z_{\text{Chern-Simons}}|=\frac{2^{N(N-1)/2}}{ k^{N/2}}
\prod_{m=1}^{N-1}|\sin^{N-m}\left(\frac{\pi m}{k}\right)|=\frac{1}{ k^{N/2}}\prod_{1\leq j<l\leq N}\big|2\sin\left(\frac{\pi(l-j)}{k}\right)\big|,
\end{align}
which is same as the one for the bosonic Chern-Simons theory 
with the level $k-N$.
This result is same for the theory on the ellipsoid as expected from the fact that the Chern-Simons theory is topological.
The v.e.v of the (SUSY) Wilson loop of the fundamental representation was also computed as \cite{KWY1}
\begin{align}
\label{ChernSimonsW}
 \langle W \rangle \sim
\prod_{1\leq j<l \leq N}
{\sin \left( {\pi (l-j+\delta_{1,j}-\delta_{1,l}) \over k}  \right)   \over \sin \left( {\pi (l-j) \over k}  \right)}
={\sin  ({\pi N \over k})  \over 
\sin \left(\frac{\pi }{k}\right)}
\end{align}
where we neglected the phase factor
which is related to the flaming dependence.

For $N > k$, this partition function (\ref{ChernSimons1}) becomes zero and this is related to the spontaneous symmetry breaking of SUSY Chern-Simons theory, which is scale invariant.
Although the partition function is diverging, the v.e.v of the Wilson loop is not diverging\footnote{
There may be Nambu-Goldstone 
fermion zero modes which make the partition function vanishes.
It is expected that the Wilson loop does not include these zero modes and the contributions of the zero modes are canceled.
}
and ${\cal O} (N^0)$ for the large $N$ limit with finite $k(\neq 1)$, which is same as for the large $N$ solution\footnote{
For $k \rightarrow 1$, we find $\langle W \rangle \sim N$.
Thus, for $k=1$, the large $N$ solution does not seem 
to correspond to the pure Chern-Simons theory.
}.
On the other hand, the free energy is divergent for 
$N > k$, which is different from the large $N$ solution for the confinement phase.
However, below we will argue that in the large $N$ expansion
the free energy $-\log |Z_{\text{Chern-Simons}}|$ is
consistent with the one for the large $N$ solution. Thus we expect that the large $N$ solution corresponds to
the SUSY breaking phase.

\begin{figure}[h!]
 \begin{minipage}{0.5\hsize}
 \begin{center}
\includegraphics[width=8cm]{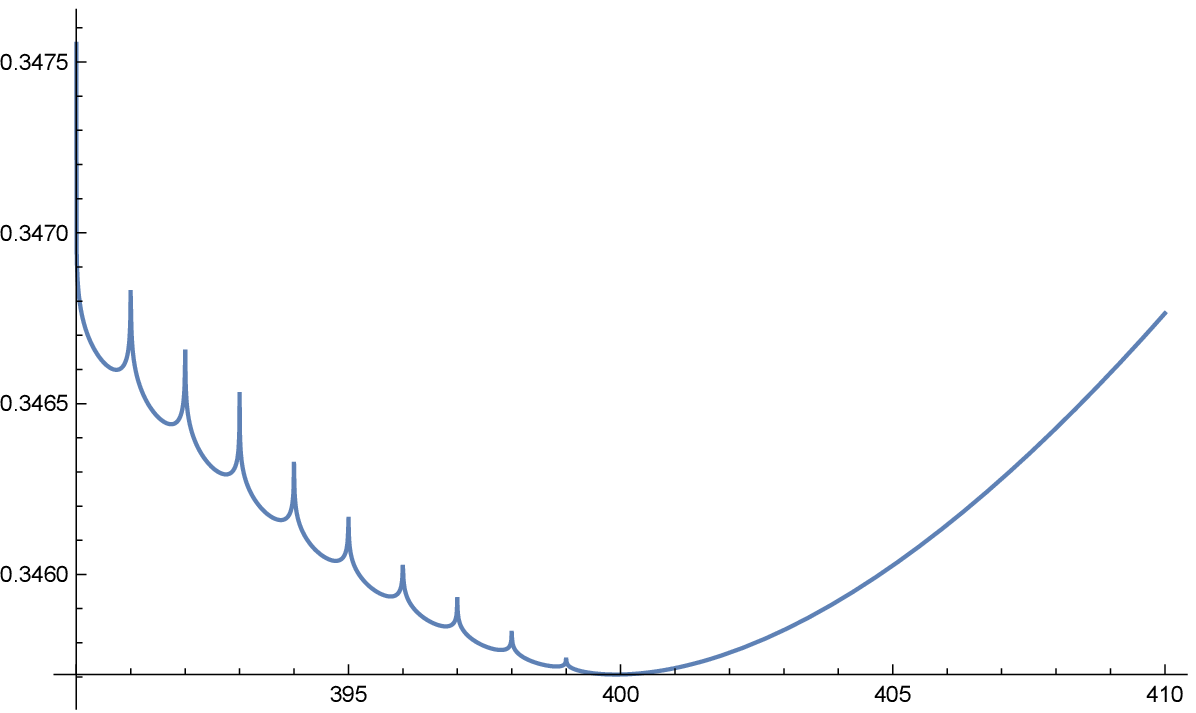}
 \label{exactCS1}
 \end{center}
 \end{minipage}
  \begin{minipage}{0.5\hsize}
 \begin{center}
 \includegraphics[width=8cm]{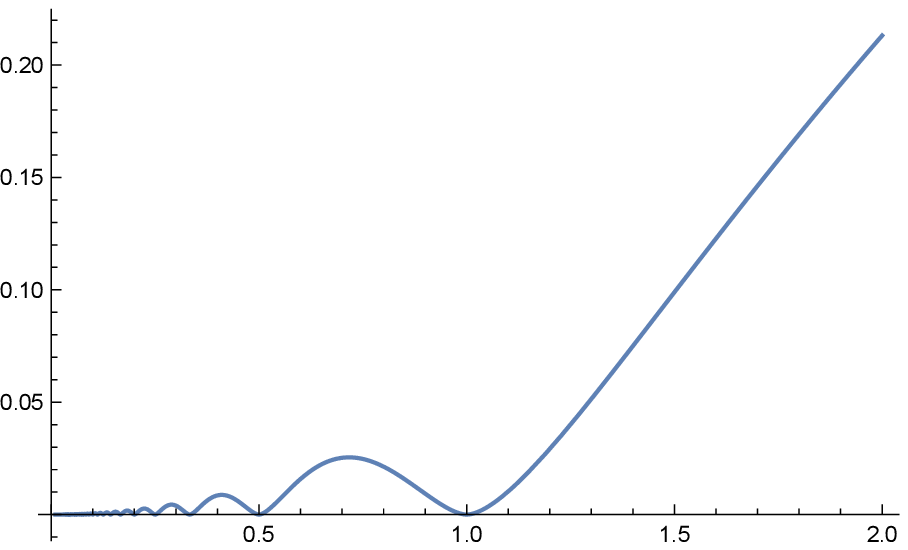}
 \end{center}
 \end{minipage}
\caption{The left figure shows the numerical plot of $-\log\big| Z_{\text{Chern-Simons}}\big|$
divided by $N^2$ with $N=400$. The horizontal axis corresponds to $k\in [390,410]$ as real value. The function is diverging when $k<N$ and $k$ is integer. 
The right figure shows \eqref{freeCS2} as a function of $t$. 
The zeros only appear in $t \leq 1$ region.
}
 \label{exactCS2}
 \end{figure}

Now we will evaluate the \eqref{ChernSimons1} in the large $N$ limit and compare it to the saddle point approximation by the large $N$ solution. 
We take logarithm of \eqref{ChernSimons1} and take continuous limit by the following replacements of the discrete index $m$ and 
the summation:
\begin{align}
\frac{m}{N}\rightarrow x\in[0,1],\quad \frac{1}{N}\sum_{m=1}^{N-1}\rightarrow \int_{0}^{1} dx.
\end{align}
Then, 
the large $N$ leading part of the logarithm of \eqref{ChernSimons1} is given by 
\begin{align}
\frac{F(t)}{N^2}\equiv -\frac{\log\big|Z_{\text{Chern-Simons}}\big|}{N^2} \sim &-\int_{0\leq y<x\leq 1} dxdy\log\big|2\sin \frac{\pi (x-y)}{t}\big|\\
\label{freeCS2}
=&-\int_{0}^{1}dx(1-x)\log\big|2\sin \frac{\pi x}{t}\big|,
\end{align}
where $N$ is infinite while keeping $\frac{k}{N}=t$ finite \footnote{
In \cite{MN,Suyama1} the authors discuss this free energy of the pure Chern-Simons theory in the 't Hooft limit. They argue that when $\lambda<1$ the integrand of the free energy has a logarithmic branch cut on a part of the integral interval and the integral is ill-defined. However, in this paper, we only consider the absolute value of the partition function and then its singular properties does not change. The integrand of \eqref{freeCS2} is well-defined even when $\lambda<1$. $t$ is a inverse 't Hooft coupling. 
}. 
In the first line of the R.H.S is same as \eqref{lf}
except $1/t$ factor.
We can find that there are infinite zeros of \eqref{freeCS2} as a function of $t$ at $t=\frac{1}{n},\ n\in \mathbb{Z}$ and no singularities for $t>0$. 
 The behavior of \eqref{freeCS2} is shown in the figure \ref{exactCS2} and we can see that 
$\frac{F(t)}{N^2} \rightarrow 0$ in $t \rightarrow 0$ limit.
This is consistent with the large $N$ solution with finite $k$.
Note that in this large $N$ leading analysis
it may be impossible to recover the
singular behavior where $k$ is an integer.

Finally, we will comment on the numerical results on the 
large $N$ solution in the `t Hooft limit.
In the region $k\ll N$ the solution is indeed the confinement solution we discussed above since the Chern-Simons term can be ignored. 
In the region $k\gg N$ we found the solution also in \cite{NST2} and the distribution of $a_j$ lies on a line in the complex plane. The numerical result suggests that as $k$ decreases and goes beyond to $k=N$, the imaginary part of the saddle point solution tends to become a double valued function as a function of the real part. The confinement solution \eqref{s1} seems to be also a double valued function including the real part of the confinement solution, which is $\mathcal{O}(\frac{1}{N})$ and ignored in this paper 
\footnote{
the real part of the solution was considered in the appendix of \cite{NST2}.
}. 
These similarities also suggest that the confinement solution
corresponds to the SUSY breaking phase.

\subsubsection{${\cal N}=2$ SUSY gauge theory with fundamental matter fields}

Here we consider the $N_f$ non-chiral pair of chiral multiplets $(Q_{a},\tilde{Q}^{a})$ in the fundamental 
and the anti-fundamental representations of the gauge group $G$ and $a$ is a flavor index which runs from 1 to $N_{f}$. The total flavor symmetry is U($N_{f}$)$\times $U($N_{f}$) and we will introduce the mass $m_a$ for the $a$-th flavor by gauging U(1)$^{N_f}$ part of flavor symmetry as usual \cite{Hosomichi}
\footnote{
When we take the gauge group G=U($N$) we take the overall U(1) symmetry of U($N$) to cancel one of the flavor mass. When the Chern-Simons level and FI parameter is vanishing the matrix model always have the contribution of one massless hypermultiplet. We do not consider such a case.
}.
For this theory, the charge can be screened and we do not expect the confinement of the charges. Furthermore, the center symmetry $\mathbb{Z}_{N}$ does not exist. 
For this theory, we have
\begin{align}
S(a) =& i \pi k \sum_{i=1}^N a_i^2 - 2 \pi i \zeta  \sum_{i=1}^N a_i
-\frac{1}{2} \sum_{i,j=1, \,\, i>j}^N 
\left( \log 4 \sinh (\pi b (a_i-a_j)+\log 4 \sinh (\pi b^{-1} (a_i-a_j)
\right) \CR 
& -  
\sum_{a=1}^{N_f} \sum_{i=1}^N 
\left(
\log s_b (\alpha+m_a-a_i )
+\log s_b (\alpha-m_a+a_i )
\right) + N\log N ,
\label{sf1}
\end{align}
where 
\begin{align}
 \alpha=\frac{iQ(1-r)}{2}.
\end{align} 
In the large $N$ limit, 
the matter parts do not contribute to the
saddle point equation for $N_f = {\cal O}(N^0)$.
Thus, the previous solution (\ref{s1}) without matter fields is valid 
for this case
although the ${\cal O}(N)$ part of the free energy $F$
is changed.

For $N_f={\cal O}(N)$,
the saddle point equation has the following extra terms:
\begin{align}
-\sum_{a=1}^{N_f}
\frac{\partial }{\partial a_{i}}
\left(
\log s_b (\alpha+m_{a}-a_{i} )
+\log s_b (\alpha-m_{a}+a_{i} )
\right), 
\end{align}
which will make the solution completely different,
which may correspond to the general fact that 
definitions of the confinement phase are ambiguous\footnote{
The center symmetry is explicitly broken by the matter fields
and the Wilson loop will not obey the area law by the pair creations.} 
for the theory with the matter fields of the fundamental representations.
However, for the large mass limit $e^{-2 \pi b^{\pm 1} |m_a|} \ll 1$,
the extra terms in the action can be approximated to 
\begin{align}
&-i\ \frac{\pi}{2} \sum_{a=1}^{N_f} \sum_{i=1}^N 
{\rm sign} (m_a) \left(
(\alpha+m_a-(a_i) )^2
-(\alpha-m_a+(a_i) )^2
\right)
\\ \nonumber 
=&-2 \pi i \alpha \sum_{a=1}^{N_f} \sum_{i=1}^N 
{\rm sgn} (m_a) 
(m_a-a_i),
\end{align}
thus we have
\begin{align}
S(a) \approx& 
-\frac{1}{2} \sum_{i,j=1, \,\, i>j}^N 
\left( \log 4 \sinh (\pi b (a_i-a_j))+\log 4 \sinh (\pi b^{-1} (a_i-a_j))
\right) + N \log N \CR 
&   
+i \pi k \sum_{i=1}^N 
\left(
a_i -\frac{1}{k} \tilde{\zeta}
\right)^2
-2 \pi i \alpha N \sum_{a=1}^{N_f} |m_a|
- i \frac{\pi}{k} \tilde{\zeta}^2,
\label{sf2}
\end{align}
where
\begin{align}
 \tilde{\zeta} = \zeta-\alpha \sum_{a=1}^{N_f} {\rm sgn}(m_a).
\end{align}
 This action is the same form as the one without the matter fields.
Thus, by shifting the U(1) part of $a_i$ in
the solution \eqref{s1},
the saddle point solution is obtained as 
\begin{align}
 a_j=i \left( \frac{j}{N} -c
\right) \sqrt{p q}M
+\frac{\tilde{\zeta}}{k},
\end{align}
and the free energy is
\begin{align}
F = \pi Q (1-r) N \sum_{a=1}^{N_f} |m_a|
- i \frac{\pi}{k} \tilde{\zeta}^2
  + {\cal O}(N).
\end{align}
The Wilson loop vanishes also because 
the shift changes the overall phase only.
Note that for the SU($N$) gauge theory,
(\ref{sf2}) with $\tilde{\zeta}=0$ is correct.

Without the vector multiplets,
in the large mass limit we have 
$F = \pi Q (1-r) N \sum_{a=1}^{N_f}|m_a|$.
Naively considering, by choosing $\tilde{\zeta}=0$,
the result of 
the pure ${\cal N}=2$ super Yang-Mills Chern-Simons theory
is recovered by subtracting the contribution of the decoupled massive
matter fields.

As discussed in the appendix for the adjoint matter fields, 
we need to choose the coupling constants
to realize the solution for the confinement phase 
for this theory with fundamental matter field
although we do not explicitly do this.

\subsubsection{${\cal N}=2$ SUSY gauge theory with adjoint matter fields}

Next we consider the $N_a$ chiral multiplets of the adjoint representation
of the gauge group $G$\footnote{
When G is SU(N) case, the matrix model of the gauge theory with gauge group G with $N_{a}\geq 2$ converges. 
However, $G=$U(N) case, the matrix model with any $N_{a}$ diverges because the integrand depends only on the difference of the integral valuables as $a_{i}-a_{j}$.
Generally, the condition that the determinant is 1 makes the matrix model to converge depending on the $N_{a}$.}. 
We will introduce the mass $m_a$ for the adjoint chiral multiplets.
Then, we have
\begin{align}
S(a) =& i \pi k \sum_{i=1}^N a_i^2 - 2 \pi i \zeta  \sum_{i=1}^N a_i
-\frac{1}{2} \sum_{i,j=1, \,\, i>j}^N 
\left( \log 4 \sinh (\pi b (a_i-a_j)+\log 4 \sinh (\pi b^{-1} (a_i-a_j)
\right) \CR 
&
+ N \log N 
-  \sum_{a=1}^{N_a}
\sum_{i,j=1
}^N 
\log s_b (\alpha+m_a-(a_i-a_j) ).
\label{sa2}
\end{align}

Comparing with the pure Chern-Simons Yang-Mills case, 
the additional terms in the saddle point equations are
\begin{align}
&\sum_{a=1}^{N_a}
\sum_{j \neq i}^N 
\frac{\partial }{\partial a_i}
\left(
\log s_b (\alpha+m_a-(a_i-a_j) ) 
-\log s_b (\alpha+m_a-(a_j-a_i) ) 
\right)
\nonumber \\
\end{align}
where $x=\frac{j}{N}$.
 We will expand $\log s_\alpha$ using (\ref{ex1}) and (\ref{ex2}).

Then, the matter contributions to the action is
\begin{align}
&&-i \frac{\pi}{2} \sum_{a=1}^{N_a} \sum_{i,j=1}^N 
{\rm sgn} (m_a) \left(
\left(\alpha+m_a-(a_i-a_j) \right)^2
+\frac{(Q^2-2)}{12}-\frac{2}{\pi} H \left(
{\rm sgn} (m_a)(\alpha+m_a-   (a_i-a_j)) \right)
\right), 
\end{align}
where 
\begin{align}
H(z)= 
\sum_{l=1}^{\infty} \frac{(-1)^{l-1}}{l} 
\left(
\frac{e^{- 2\pi l bz}}{2 \sin(\pi l b^2)} +
\frac{e^{- 2\pi l z/b}}{2 \sin(\pi l b^{-2})} 
\right).
\end{align}
Now we assume 
\begin{align}
 \sum_{a=1}^{N_a} {\rm sgn} (m_a) =0,
\end{align}
to solve the saddle point equations.

Below, we will show that
\begin{align}
 a_j=2 i \left( \frac{j}{N} -c
\right) \sqrt{p q}M,
\label{s2}
\end{align}
is a solution of the saddle point equation
for the theory with the adjoint matter fields\footnote{
There is an additional factor 2 in (\ref{s2}) compared with
the pure Yang-Mills case.
This is because $H(x)$ contains
$e^{\pi b^{\pm1} x}$ instead of $e^{2 \pi b^{\pm1} x}$
for $\sinh^2 (\pi b^{\pm 1} x)$. 
If $(1+b^{\pm 2})(1-r)=m_{\pm}$ where $m_{\pm} \in {\mathbb Z}$,
we can show that this factor 2 can be dropped,
although these mean $b=2r=m_{\pm}=1$ for $r>0$.}.
We see that the terms including $H^{\prime}(x)$ vanish
because 
$H(x)$ is  
the periodic function with the period $2 i\sqrt{pq} $, i.e. $H(x+2 i\sqrt{pq} )=H(x)$,
which can be seen from $\sqrt{pq} =q  b =p /b$, and then,
\begin{align}
\label{priod}
\sum_{j\neq i}^{N}H^{\prime}\left(\text{sgn}(m_{a})\left(\alpha+m_a\pm i\frac{\sqrt{pq}M}{N} \left(i-j\right)\right)\right)=
\sum_{\ell =1}^{N-1}H^{\prime}\left(\text{sgn}(m_{a})\left(\alpha+m_a\pm i\frac{\sqrt{pq}M}{N} \ell\right)\right).
\end{align}
We can also see that the remaining terms are canceled each others:
\begin{align}
\sum_{a=1}^{N_a}
{\rm sgn}(m_a) 
\left(
\alpha+m_a+2 i\frac{\sqrt{pq}M}{N}\sum_{j\neq i}^{N}\left(i-j\right)
-(\alpha+m_a)+2 i\frac{\sqrt{pq}M}{N} \sum_{j\neq i}^{N}\left(i-j\right)
\right)
=0.
\end{align}
Therefore (\ref{s2}) is the large $N$ saddle point solution.
The Wilson loop vanishes in the large $N$ limit 
as for the pure Chern-Simons Yang-Mills case.
We can also compute the free energy and 
find $F=0 \cdot N^2+{\cal O} (N)$ where we used the periodicity 
of the function $H(z)$ like as \eqref{priod}.

For this theory with fundamental matter fields,
there is another large $N$ solution as 
discussed in the appendix.
To realize the confinement phase,
we need to choose the coupling constants
as we will see in the appendix.

\section{Mass deformed ABJM theory on $S^3_b$}

The mass deformed ABJM theory \cite{HLLLP,GRVV}, 
which is obtained from the ABJM theory by a mass deformation, has some interesting properties.
This mass deformed ABJM  theory preserves $\mathcal{N}=6$ SUSY and it is known that the vacuum solutions compose the Fuzzy three sphere \cite{GRVV} and corresponds to the M2-M5 branes system, which is an analog of the Myers effects \cite{M}.

In \cite{NST1} and \cite{NST2}, we considered the large $N$ limit of the round sphere partition function of
the mass deformed ABJM theory and found that the large $N$ solution of the saddle point equation, for which the free energy is proportional to $N^{\frac32}$, is valid only when $\frac{\zeta}{k}<\frac{1}{4}$, where $\zeta$ is the  FI parameter and $k$ is Chern-Simons level. 
If the mass is larger than this critical value, it was argued that the SUSY breaking occurs in \cite{HNST}.
In this phase, the large $N$ solution of the confinement phase
can be realized.
Indeed, for the bi-fundamental matter fields as in the ABJM theory, 
we will follow the previous 
discussion on the adjoint matter field and easily find 
the large $N$ solution obtained there on the ellipsoid 
is the solution for the bi-fundamental matter fields.
Furthermore, there are some extensions of this model
to M2-branes in other backgrounds.
They have generally gauge theories with a product group
and bi-fundamental matter fields, then
the large $N$ solution of the confinement phase can be relevant.
Thus, these models are interesting candidates for which the large $N$ solution of the confinement phase is relevant.

One thing we need to consider for the mass deformed ABJM theory
is the critical value of the mass for the theory on the ellipsoid.
In order to do this, we study the large $N$ solution, for which the free energy is proportional to $N^{\frac32}$,  on the ellipsoid. 
In this section, we will consider the partition function of the mass deformed ABJM theory on ellipsoid $S^{3}_{b}$ in the large $N$ limit. 
We will see that how the critical value for the mass parameter is modified by the ellipsoid parameter. 

For the mass deformed ABJM, the action is 
\begin{align}
S(a)=S_0+S_1+ 2 N \log N,
\end{align}
where 
\begin{align}
S_0=& i \pi k \sum_{i=1}^N (a_i^2-\tilde{a}_i^2) - 2 \pi i \zeta  \sum_{i=1}^N (a_i+\tilde{a}_i),
\end{align}
and
\begin{align}
S_1=
& -\frac{1}{2} \sum_{i,j=1, \,\, i>j}^N 
\left( \log 4 \sinh^2 (\pi b (a_i-a_j)+\log 4 \sinh^2 (\pi b^{-1} (a_i-a_j)
\right) \CR 
& -\frac{1}{2} \sum_{i,j=1, \,\, i>j}^N 
\left( \log 4 \sinh^2 (\pi b (\tilde{a}_i-\tilde{a}_j)+\log 4 \sinh^2 (\pi b^{-1} (\tilde{a}_i-\tilde{a}_j)
\right) \CR 
& -  
\sum_{i,j=1}^N \left( \log D_b(a_i-\tilde{a}_j) + \log D_b(a_i-\tilde{a}_j) 
\right).
\label{amabjm}
\end{align}

For the large $N$ solution which has a gravity dual, 
we use the continuous notation $a_i \rightarrow a(s)$ with $s \sim \frac{i}{N} +{\rm const.}$
and take the following form \cite{NST2}:
\begin{align}
a(s) = \sqrt{N} z_1(s) +z_2(s), \CR
\tilde{a}(s) = \sqrt{N} z_1(s) -z_2(s), 
\end{align}
where $z_{1,2}$ are independent arbitrary complex valued functions of $s$.
For $S_0$ which is the classical part of $S(a)$, we can easily evaluate the 
large $N$ leading contribution as 
\begin{align}
 i \pi k \sum_{i=1}^N (a_i^2-\tilde{a}_i^2) - 2 \pi i \zeta  \sum_{i=1}^N (a_i+\tilde{a}_i)
\approx 4 \pi N^{\frac32} \int ds (i k z_1 z_2 - i \zeta z_1),
\end{align}
For the remaining 1-loop part $S_1$,
we define 
\begin{align}
z(s)=\sqrt{N} (z_1(s)-z_1(s')), \,\,\,\,\,\, 
w_{\pm} (s)= z_2(s)\pm z_2(s'),
\end{align}
with a fixed $s'$.
Then, we find
\begin{align}
S_1= &- \frac12 N^2 \int_0^1 d s' \int _{s'}^1 ds
[ \log 4 \sinh^2 (\pi b (z(s)+w_-(s)))
+ \log 4 \sinh^2 (\pi b (z(s)-w_-(s))) \CR
& \hspace{3cm}
+\log 4 \sinh^2 (\pi b^{-1} (z(s)+w_-(s)))
+ \log 4 \sinh^2 (\pi b^{-1} (z(s)-w_-(s)))
] \CR
& - N^2 \int_0^1 d s' \int _{s'}^1 ds 
[
\log D_b(z(s)+w_+(s)) +\log D_b(-z(s)+w_+(s))  \CR
&   \hspace{3cm} +\log D_b(z(s)+w_+(s)) +\log D_b(-z(s)+w_+(s)) 
],  \CR
\equiv &- N^2 \int_0^1 d s' \int _{s'}^1 ds \log Y.
\end{align}
where we defined $Y$ as 
\begin{align}
Y & \equiv   V_{+} \cdot V_{-}\cdot D_b(z+w_+) \cdot D_b(z+w_+) \cdot D_b(z-w_+) \cdot D_b(z-w_+),
\end{align}
where the vector multiplet 1-loop function is rewritten as
\begin{align}
\log V_{\pm}(s)=&\frac{1}{2}\log \left(4\sinh^2\pi b\left(z(s)\pm w_{-}(s)\right)4\sinh^2\pi b^{-1}\left(z(s)\pm w_{-}(s)\right)\right).
\end{align}
In the large $N$ limit the leading part of $\log V_{\pm}(s)$ is evaluated as 
\begin{align}
\log V_{\pm}(s)=& \pi Q(z(s)\pm w_{-}(s))+\log\left[\left(1-e^{-2\pi b(z(s)\pm w_{-}(s))}\right)\left(1-e^{-2\pi b^{-1}(z(s)\pm w_{-}(s))}\right)\right].
\end{align}
The first term is the leading term which is order $\mathcal{O}(\sqrt{N})$ and this term cancels with the one-loop part of the bi-fundamental of hypermultiplets. The one-loop part of the bi-fundamental hypermultiplets can be written for suitable form when $x$ has a positive real part as
\begin{align}
\label{asymd}
\log D_{b}(x)=&-\frac{\pi Q x}{2}+\int_{\mathbb{R}+i0}\frac{dt}{4t}\frac{\sinh\left(\frac{Qt}{2}\right)e^{2ixt}}{\sinh bt\sinh b^{-1}t}+\int_{\mathbb{R}-i0}\frac{dt}{4t}\frac{\sinh\left(\frac{Qt}{2}\right)e^{-2ixt}}{\sinh bt\sinh b^{-1}t}.
\end{align}
This form is followed from the integral representation of the $D_{b}$ \eqref{integral}.
The integral contour $\mathbb{R} \pm i0$ in \eqref {asymd} means starlight line from $-\infty$ to $\infty$ which avoids from  the origin by moving slightly to upper/lower half complex plane respectably. In order to get the \eqref{asymd}  we expand $\cos(2xt)$ as exponential and change integral contour $\mathbb{R}+i0$ to $\mathbb{R}-i0$ and pick up residue at the origin. The residue contribution is given by the first linear term and the second and third terms are $\mathcal{O}(e^{-\sqrt{N}})$ in this case and can be ignored except when $\text{Re}(z)=0$.

The leading order of $\log Y$ seems to be $\mathcal{O}(\sqrt{N})$, however, the leading parts cancel totally between the vector multiplets and bi-fundamental hypermultiplets part as follows:
\begin{align}
&\int^{\frac{1}{2}}_{-\frac{1}{2}}ds^{\prime}\int^{\frac{1}{2}}_{s^{\prime}} ds \log Y \sim \nonumber \\
&\pi Q\int^{\frac{1}{2}}_{-\frac{1}{2}}ds^{\prime}\int^{\frac{1}{2}}_{s^{\prime}} ds \left(z(s)+w_{-}(s)+z(s)-w_{-}(s))-(z(s)+w_{+}(s)+z(s)-w_{+}(s)\right)=0.
\end{align}

Thus, the
only region near $|{\rm Re}( z)| =0$, i.e. $s=s'$, in the integration of $s'$
gives the large $N$ true leading contribution of $S_1$.
Furthermore as discussed in \cite{NST2}, 
the saddle point solution $z_1(s)$
should be a monotonically increasing function of $s$.
Assuming this, we can evaluate the large $N$ leading contribution of $S_1$ by the following relation:
\begin{align}
\int_{s^{\prime}}^{\frac{1}{2}}ds \log \left(V_{+}(s)V_{-}(s)e^{-2\pi Q z(s)}\right)\sim & \frac{1}{\sqrt{N}\dot{z}_{1}(s)}\int_{C_{0}} dz \log(1-e^{-2\pi bz })^2(1-e^{-2\pi b^{-1} z})^2\nonumber \\
=&-\frac{1}{\sqrt{N}\dot{z}_{1}(s^{\prime})}\frac{\pi}{6}Q,
\end{align}
\begin{align}
\label{hyperappro}
&\int_{s^{\prime}}^{\frac{1}{2}} ds \log\left(e^{\pi Q z(s)}\left(D_{b}(z(s)+w_{+}(s))D_{b}\left(z(s)-w_{+}(s)\right)\right)\right) \nonumber \\
\sim& \left(\int_{C_{+}}+\int_{C_{-}}\right) dz \frac{1}{\sqrt{N}\dot{z}_{1}(s^{\prime})}\log\left(e^{\frac{\pi Q z}{2}}D_{b}(z)\right) \nonumber \\
=&\frac{1}{\sqrt{N}\dot{z}_{1}(s^{\prime})}\left(2\int_{0}^{\infty}+\int_{2z_{2}(s^{\prime})}^{0}+\int_{-2z_{2}(s^{\prime})}^{0}\right)dz\log\left(e^{\frac{\pi Q z}{2}}D_{b}(z)\right)\nonumber \\
=&\frac{1}{\sqrt{N}\dot{z}_{1}(s^{\prime})}\left(2\int_{0}^{\infty}dz\left(\int_{\mathbb{R}+i0}\frac{dt}{4t}\frac{\sinh\left(\frac{Qt}{2}\right)e^{2izt}}{\sinh bt\sinh b^{-1}t}+\int_{\mathbb{R}-i0}\frac{dt}{4t}\frac{\sinh\left(\frac{Qt}{2}\right)e^{-2izt}}{\sinh bt\sinh b^{-1}t}\right) \right.\nonumber \\ 
 &\quad\quad\quad\quad\quad+\left. \int_{2 z_{2}(s)}^{0}dz\log\left(e^{\pi Q z}\right)\right) \nonumber \\
=&-\frac{\pi Q}{\sqrt{N}\dot{z}_{1}(s^{\prime})}\left(\frac{1}{24}\left(b^2+b^{-2}-\frac{Q^2}{4}\right)+2\left(z_{2}(s)\right)^2\right).
\end{align}
$z(s)$ is varied vary largely even when $s$ is slightly change around $s=s^{\prime}$ because $z(s)$ is proportional to $\sqrt{N}$. Then the $C_{0}$ is regarded as the half-starlight line from the origin  towards $z(\frac{1}{2})$. The integral contours  $C_{\pm}$ also are regarded as the half-starlight line from $\pm2z_{2}(s)$ towards $z(\frac{1}{2})$ respectably. $\dot{z}_{1}(s)$ can be replaced by  $\dot{z}_{1}(s^{\prime})$  and we can take $w_{-}=0,\  w_{+}=2z_{2}(s^{\prime})$. In the second line of \eqref{hyperappro} we change the integral contour without crossing the poles of $\log D_{b}$ assuming the condition is satisfied 
\begin{align}
-\frac{Q}{8}<\text{Im}\left(z_{2}(s)\right)-\text{Re}\left(z_{2}(s)\right))\frac{\text{Im}(\dot{z}_{1}(s))}{\text{Re}\left(\dot{z}_{1}(s)\right)}<\frac{Q}{8}.
\end{align} 
The first and second terms in the third line of \eqref{hyperappro} are evaluated as by commuting the integrals
\begin{align}
\left(-\int_{\mathbb{R}+i0}\frac{dt}{8it^2}\frac{\sinh\left(\frac{Qt}{2}\right)}{\sinh bt\sinh b^{-1}t}+\int_{\mathbb{R}-i0}\frac{dt}{8it^2}\frac{\sinh\left(\frac{Qt}{2}\right)}{\sinh bt\sinh b^{-1}t}\right)=&\oint_{t=0}\frac{dt}{8it^2}\frac{\sinh\left(\frac{Qt}{2}\right)}{\sinh bt\sinh b^{-1}t}\nonumber \\
=&\frac{\pi Q}{48}\left(\frac{Q^2}{4}-(b^2+b^{-2})\right).
\end{align}
Then the leading part of one-loop action is evaluated as 
\begin{align}
S_{1}=&-\pi QN^{\frac{3}{2}}\int_{-\frac{1}{2}}^{\frac{1}{2}}\frac{ds}{\dot{z}_{1}(s)}\left(\frac{1}{12}\left(b^2+b^{-2}+2-\frac{Q^2}{4}\right)+4(z_{2}(s))^2\right)\nonumber \\
=&-4\pi QN^{\frac{3}{2}}\int_{-\frac{1}{2}}^{\frac{1}{2}}\frac{ds}{\dot{z}_{1}(s)}\left(\frac{Q^2}{64}+(z_{2}(s))^2\right).
\end{align}
Consequently, the leading part of  total action of the mass deformed ABJM theory is
\begin{align}
\label{finalformabjm}
S=&4\pi N^{\frac{3}{2}} \int_{-\frac{1}{2}}^{\frac{1}{2}}ds\left(ikz_{1}(s)z_{2}(s)-i\zeta z_{1}(s)-\frac{Q}{\dot{z}_{1}(s)}\left(\frac{Q^2}{64}+(z_{2}(s))^2\right)\right)\nonumber \\
&=\pi Q^2N^{\frac{3}{2}}\int_{-\frac{1}{2}}^{\frac{1}{2}}ds \left(ik\tilde{z}_{1}(s)\tilde{z}_{2}(s)-i\tilde{\zeta}\tilde{z}_{1}(s)-\frac{2}{\dot{\tilde{z}}_{1}(s)}\left(\frac{1}{16}+(\tilde{z}_{2}(s))^2\right)\right),
\end{align}
where in the last line we rescale variables as following:
\begin{align}
z_{1,2}(s)=\frac{Q}{2}\tilde{z}_{1,2}(s)\ ,\quad \zeta=\frac{Q}{2}\tilde{\zeta}.
\end{align}
Then the action \eqref{finalformabjm} is same as that of mass deformed ABJM on round three-sphere which we obtained in \cite{NST2} up to overall factor. The solution of the saddle point equation is same in terms of $\tilde{z}_{1,2}(s)$. The interesting point is that the bound where the saddle point solution exists depends on ellipsoid parameter $Q$ like as
\begin{align}
\frac{\tilde{\zeta}}{k}<\frac{1}{4} \Leftrightarrow \frac{\zeta}{k}<\frac{Q}{8}.
\label{cv}
\end{align}
The free energy is given by
\begin{align}
F=\frac{\pi Q^2\sqrt{2k} N^{\frac{3}{2}}}{12}\left(1+\left(\frac{4\tilde{\zeta}}{k}\right)^2\right)=\frac{\pi Q^2\sqrt{2k} N^{\frac{3}{2}} }{12}\left(1+\left(\frac{8\zeta}{Qk}\right)^2\right),
\end{align}
in the large $N$ limit\footnote{When we take $\zeta =0$ this result coincides with that obtained in \cite{ID}.}.
We expect that the large $N$ solution considered in this section 
is valid for (\ref{cv}) and 
the SUSY breaking occurs
if the mass is above this region.
In particular, in the large $b$ limit,
the critical value for $\zeta$ is $\zeta_c=k b/8$.
Thus, $l_1 \zeta_c=k/8$ is finite in this large $b$ limit
where $l_1$ is the length of $S^1$ in the limiting geometry, i.e. $S^1 \times {\mathbb R}^2$.
This indicates that the SUSY breaking phase also appears above the critical mass 
on $S^1 \times {\mathbb R}^2$.
This might be possible because of the boundary condition of the spatial infinity of $ {\mathbb R}^2$,
which should be determined by the $b \rightarrow \infty$ limit of the ellipsoid, 
may select the metastable SUSY vacuum which may be stable in the large $N$ limit. 
Note that the free energy $F$ contains the contributions from 
the outside of $\theta \sim 0$ region which is approximated by $S^1 \times {\mathbb R}^2$
and diverges in the $b \rightarrow \infty$ limit.

\section{Summary and discussion}

In this paper, we have studied the properties of the confinement solution which we discovered in \cite{NST2} and the theory which has that type of the theory.

First, we have considered the theory on $S^{3}_{b}$ for supersymmetric Wilson loops to 
    the Polyakov loop by taking $b\rightarrow \infty$ limit in the sense that $S^{3}_{b}$ become locally $S^{1} \times \mathbb{R}^{2}$ and the supersymmetric Wilson loop is wrapping on the $S^{1}$. In the large $N$ limit supersymmetric Wilson loop can be evaluated with the solution of the saddle point equation. We showed that various gauge theories have the special kind of solution. With this solution, the Wilson loop is vanishing in the large $N$ limit. We call the solution as confinement solution in the sense that this Wilson loop can be regarded as the generalized Polyakov loop. 
We expect this solution corresponds to the spontaneously SUSY breaking phase. 
One reason for this expectation is 
that this solution only valid in the region $N \gg k $. This is consistent with the fact the SUSY breaking phase may be gapped and confined \cite{Witten1}.

Then, we discussed the parameter region where this solution is valid for the mass deformed ABJM theory and other theories on $S^{3}_{b}$. In \cite{NST2} it is showed that the mass deformed ABJM theory has the large $N$ solution whose free energy is proportional to $N^{\frac{3}{2}}$ and it is valid when $\frac{\zeta}{k}<\frac{1}{4}$. 
When $\frac{\zeta}{k}\geq\frac{1}{4}$ it is expected that 
the theory is in the SUSY breaking phase in the large $N$ limit
and
the confinement solution can become dominant.
This might reflect into the fact that pure CS theory in a SUSY breaking phase appears if all massive bi-fundamental hypermultiplets decouple. In this paper, we found that the solution of the saddle point equation of the mass deformed ABJM  whose free energy is proportional to $N^{\frac{3}{2}}$ and that solution is also valid only when $\frac{\zeta}{k}<\frac{Q}{8}$. 
Thus, we expect that 
the theory is in the SUSY breaking phase and
the confinement solution can be relevant in the large $N$ limit when $\frac{\zeta}{k}\geq \frac{Q}{8}$.

One of the interesting future work is to consider the large $N$ 
solution for the confinement phase in the theory on the Seifert manifold \cite{CKW1,CKW2}. 
Indeed, we expect that this solution for the confinement phase exists for the theory on the Seifert manifold also by the following reasons.
In the 1-loop part of the twisted index of the Seifert manifold, a complex scalar, in which the real part is the holonomy $a$ along the $S^1$ fiber direction and the imaginary part is
the $\sigma$, appears.
Moreover, the 1-loop part is periodic under the constant shift of the holonomy which is regarded as the large gauge transformation and essentially equivalent to the action of the center symmetry. 
This also means that there is the periodicity under the shift of the imaginary part of 
the $\sigma$ because this also gives the same constant shift of the complex scalar, thus the large $N$ solution for the confinement phase exists as for the theory on the squashed $S^3$.
Note that if there is no non-trivial one-cycle, the holonomy $a$ does not exist, however, for the Seifert manifold with a non-trivial fundamental group, the large $N$ solution for the confinement is just the usual symmetric configuration of the Polyakov loop for the confinement phase.
This might explain our observation that the large $N$ solution for the theory on the squashed $S^3$ may exist only when the square of the squashed parameter $b$ is a rational number because the partition function for the theory on the squashed three-sphere with the rational squashing parameter is represented by the one on a Seifert manifold. It will be interesting to investigate the large $N$ solution for the theory on a Seifert manifold further and the corresponding gravity solutions \cite{TW,GK}.\footnote{
We would like to thank a referee of this paper for suggesting a relation between our solution and the theory on Seifert manifolds.}

In appendix A, we introduce an example which has the confinement solution and the other solution and investigate which is dominant in the large $N$ limit. We propose that the confinement solution can be dominant in the large $N$ limit with a specific region of parameters. It is interesting to find the confinement solution by the resolvent methods and interpolate it with the solution in the weak 't Hooft coupling limit. It is also interesting to know the condition where the confinement solution becomes dominant analytically.

\section*{Acknowledgments}

We would like to thank S. Yokoyama and  Y. Hatsuda for useful discussions. 
We would like to thank M. Honda and T. Nosaka for
reading the manuscript of this paper and valuable comments and discussion.
K.S. is supported by JSPS fellowship and Grant-Aid for JSPS Fellow No.18J11714. 
The work of S.T. was supported by JSPS KAKENHI Grant Number 17K05414.

\appendix

\section{Dominant saddle point solution of matrix models of $S^{3}$}

We have studied the large $N$ saddle point solution which 
corresponds to the confining phase.
In general, a large $N$ saddle point solution is not guaranteed to give the dominant contribution to the partition function. Namely, if there are many solutions, then, 
we need to find which solution will really give the dominant contribution to the partition function in the large $N$ limit.
In this appendix, we will focus 
on SU($N$) SUSY Chern-Simons Yang-Mills theory with $N_{f}$ adjoint massive hypermultiplets on $S^{3}$ 
and investigate whether there are other solutions of the saddle point equation and the confinement phase is realized or not. 
We will see that for the large $N$ limit with $N \gg k $ and $\frac{k m}{N}$ finite and large the confinement phase is 
expected to be realized.

\begin{figure}[h!]
 \begin{minipage}{0.5\hsize}
 \begin{center}
\includegraphics[width=8cm]{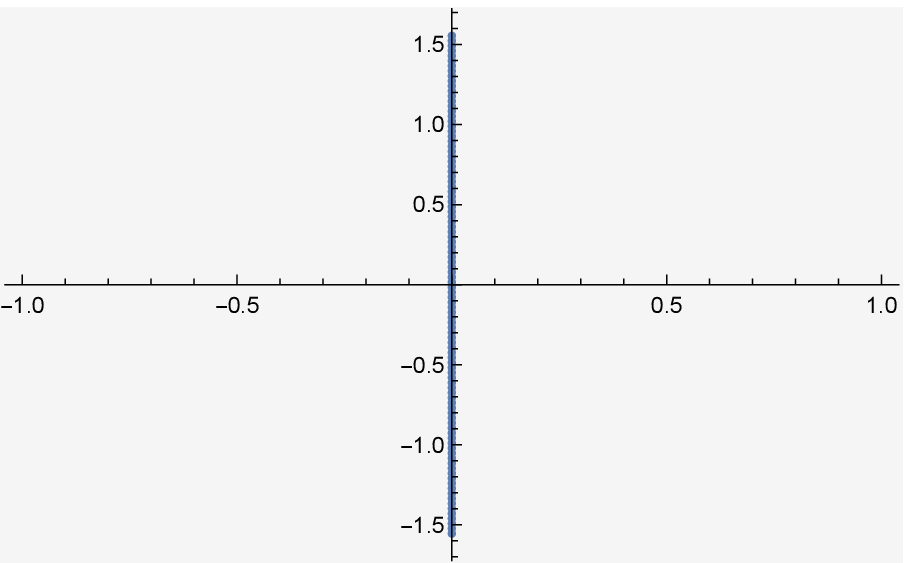}
 \end{center}
 \end{minipage}
  \begin{minipage}{0.5\hsize}
 \begin{center}
 \includegraphics[width=8cm]{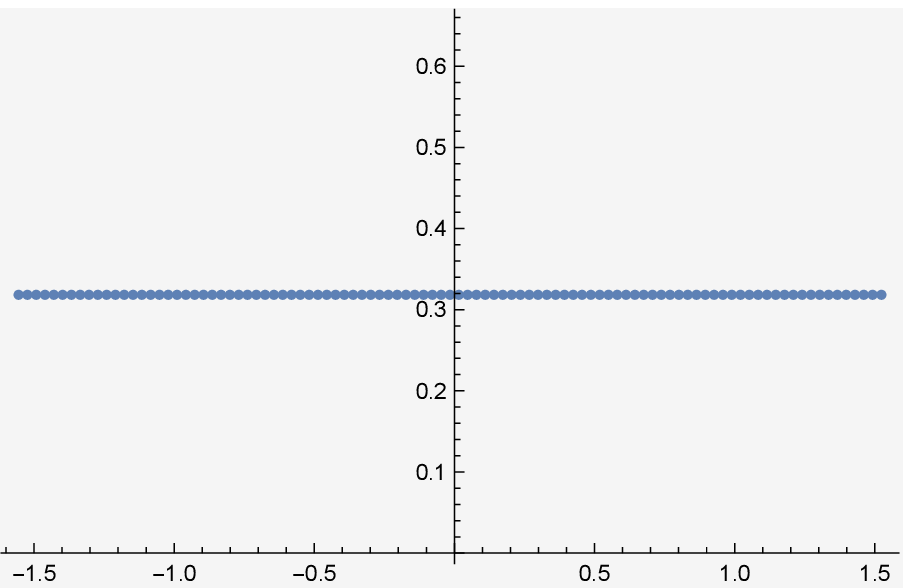}
 \end{center}
 \end{minipage}
\caption{The left figure shows the numerical solution 
which corresponds to the confinement phase
plotted on the Complex plane with the parameters $(N,N_{f},m)=(100,3,8)$. The solution actually lies on the Imaginary axis. The right one shows the density function of it. We can check that the solution does not depend on the parameters.}
  \label{figiimsol}
 \end{figure}

\begin{figure}[h!]
 \begin{minipage}{0.5\hsize}
 \begin{center}
\includegraphics[width=8cm]{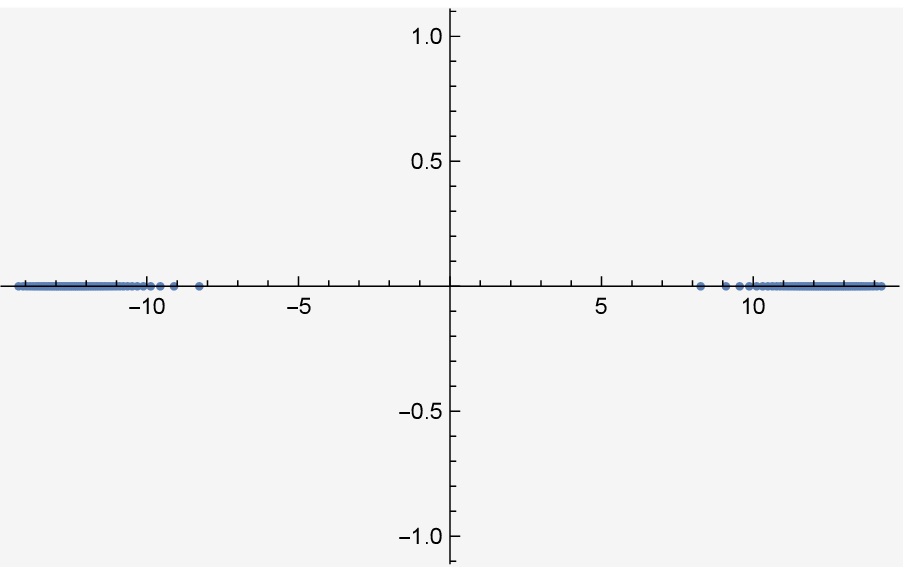}
 \end{center}
 \end{minipage}
  \begin{minipage}{0.5\hsize}
 \begin{center}
 \includegraphics[width=8cm]{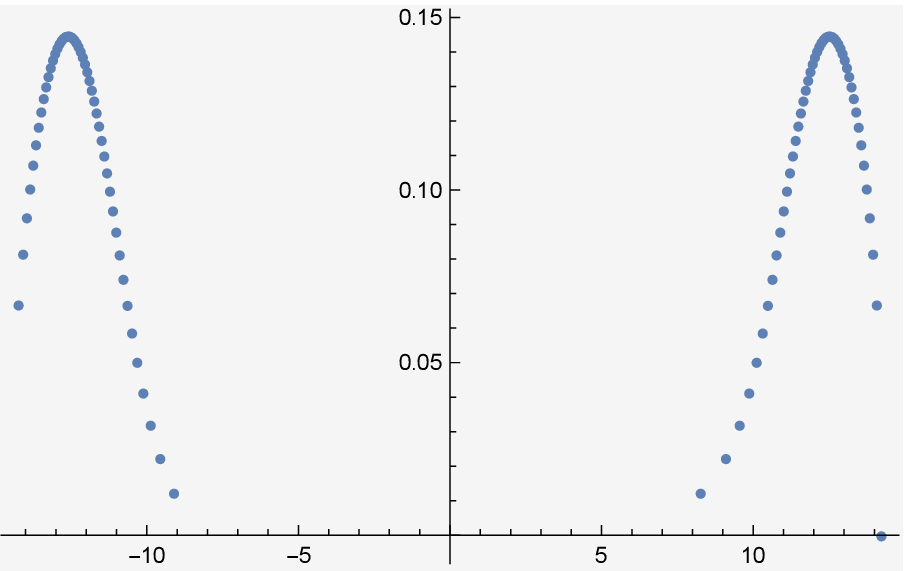}
 \end{center}
 \end{minipage}
\caption{The left figure shows the other numerical solution plotted on the Complex plane with $(N,N_{f},m)=(100,3,8)$. The solution actually lies on the real axis. The right one shows the density function of it.}
  \label{figiresol}
 \end{figure}

First, we will consider $k=0$ case.
Here the massive hypermultiplets mean that the matrix model is given by
\begin{align}
\label{SYMadj2}
Z=\frac{1}{N!}\int{d^{N}a}\delta(\sum_{i}^{N}a_{i})\prod_{i>j}^{N}\frac{4\sinh^{2}\pi(a_{i}-a_{j})}{\left(2\cosh\pi(a_{i}-a_{j}+m)2\cosh\pi(a_{i}-a_{j}-m)\right)^{N_{f}}},
\end{align} 
where $\delta(\sum_{i}^{N}a_{i})$ are reflected into the condition of SU($N$). The matrix model converges when we take $N_{f}\geq 2$. Then we take $N_{f} \geq2$ and there are no subtleties since the matrix model is well-defined. The saddle point equation of this matrix model is difficult to solve analytically. However, with the help of the numerical analysis of the saddle point equation we find two solutions of the saddle point equation, one of which is confinement solution. The other saddle point solution admits the real value solution while the confinement solution is complex. These two type of solutions are showed in figure \ref{figiimsol} and \ref{figiresol}.
The free energy for the real solution is smaller than that for the confinement one. 
Then the confinement phase can not be realized in the large $N$ limit. 
We will show some numerical values of the free energy for the two solutions in the following tables:

\begin{table}[h!]
  \begin{center}
    \begin{tabular}{|c||c|c|c|c|} \hline
              & $m_{a}=1$ &  $m_{a}=2$ & $m_{a}=5$ & $m_{a}=8$ \\ \hline
      $N_{f}=2$   &  49746& 94563 & 233023& 372313\\ 
      $N_{f}=3$  &83582 & 160235& $337392$& 632638\\
      $N_{f}=5$ &  148258 & 287081& $699586$&1125850 \\
      $N_{f}=10$  & 306535& 600494 & $ 1486380$& 2379170\\ \hline
    \end{tabular}
      \caption{The free energies correspond to the real solution.}
  \end{center}
\end{table}

\begin{table}[ht!]
  \begin{center}
    \begin{tabular}{|c||c|c|c|c|} \hline
              & $m_{a}=1$ &  $m_{a}=2$ & $m_{a}=5$ & $m_{a}=8$ \\ \hline
      $N_{f}=2$   &  61742& 123947 & 310557& 497168\\ 
      $N_{f}=3$  &92844& 186150 & $466066$& 745982\\
      $N_{f}=5$ &  155047 & 310557& $713310$&1243610 \\
      $N_{f}=10$  & 310555& 621575 & $ 1554630$&  2487680\\ \hline
    \end{tabular}
      \caption{The free energies correspond to the confinement solution.}
  \end{center}
\end{table}

Thus, the numerical results suggest that 
this theory with $k=0$ whose matrix model converges has at least the two solutions of the saddle point equation,
and the confinement solution is not dominant saddle point solution in any regions of $m$ and $N_{f}$. 
However, naively considering, in the infinite mass limit the theory become pure SU($N$) SYM since the matter fields become decoupled. 
Thus if this naive expectation might be true the confinement solution should be dominant in that limit.
However, this naive expectation is not always true. This is because the naive discussion of decoupling is based on the assumption that the theory is on the origin of the Coulomb branch. In fact, the numerical results suggest that the dominant solution of the saddle point equation does not correspond to the origin of the Coulomb branch of the flat space. 
We will discuss this 
from the numerical result with the case $N_{f}\geq 3$ in the next subsection.

\subsection{The meaning of saddle point solutions}
The matrix model is written by the integral of the eigenvalues of the Coulomb branch parameters. The saddle point solution is a specific configuration of the Coulomb branch \footnote{Exactly speaking, the solution corresponds to the real part of the classical Coulomb branch parameter.}. 
Then it is plausible to consider that in the large $N$ limit the specific point of the Coulomb branch is selected. This means that in the large $N$ limit we can argue which massive hypermultiplets are effectively massless based on the saddle point solution.

Let us consider the meaning of the real solution we introduced above. 
When we take mass sufficiently large the real saddle point configuration
split into two parts and the half of the $N$ eigenvalues are distributed around $-\frac{m}{2}$ and the others are around $\frac{m}{2}$. 
This solution means that a non-trivial point of the Coulomb branch is selected in the large $N$ limit where the theory has effective massless degrees of freedom in the deep IR of the RG flow and the theory may flow to a non-trivial interacting superconformal field theory. 
On the other hands, the confinement type solution is expected that all the massive hypermultiplets become decoupled from the theory in the infinite mass limit and the theory may be in the SUSY breaking phase.

Below we will study the split type solution and its effective theory, which is a so-called "good" theory defined in \cite{GW}. We assume that the eigenvalues which have two separated positive and negative region near $\pm \frac{m}{2}$. We assume the first $\frac{N}{2}$ eigenvalues are negative region $I_{-}$ and the other eigenvalues are positive region $I_{+}$.\footnote{
Here we also describe the indexes of the eigenvalue which is in positive region as $I_{+}$ and that of in negative as $I_{-}$.} First we consider that ${i}\in I_{-}$ and the saddle point equation: 
\begin{align}
0=\frac{\partial S(a)}{\partial a_{i}}=&2\sum_{j\in I_{-}}\coth\pi\left(a_{i}-a_{j}\right)+2\sum_{j\in I_{+}}\coth\pi\left(a_{i}-a_{j}\right)\nonumber \\
&-N_{f}\sum_{j\in I_{+}}\left(\tanh\pi(a_{i}-a_{j}+m)+\tanh\pi(a_{i}-a_{j}-m)\right)\nonumber \\
&-N_{f}\sum_{j\in I_{-}}\left(\tanh\pi(a_{i}-a_{j}+m)+\tanh\pi(a_{i}-a_{j}-m)\right)+\mu.
\end{align}
From the assumption we can take the eigenvalue as
\begin{align}
\begin{cases}
a_{i}=\frac{m}{2}+\lambda_{i}\quad i\in I_{+},\\
a_{i}=-\frac{m}{2}+\tilde{\lambda}_{i} \quad i\in I_{-}
\end{cases},
\end{align}
where $\lambda_{i}$ and $\tilde{\lambda}_{i}$ are $\mathcal{O}$(1) and the saddle point equation is evaluated as by using this assumption
\begin{align}
0=&2\sum_{j\in I_{-}}\coth\pi(\lambda_{i}-\lambda_{j})+2\sum_{j \in I_{+}}\coth\pi(-m+\lambda_{i}-\tilde{\lambda}_{i})\nonumber \\
&-N_{f}\sum_{j \in I_{-}}\left(\tanh\pi(\lambda_{i}-\lambda_{j}+m)+\tanh\pi(-m+\lambda_{i}-\lambda_{j})\right)\nonumber \\
&-N_{f}\sum_{j \in I_{+}}\left(\tanh\pi(\lambda_{i}-\tilde{\lambda}_{j})+\tanh\pi\left(\lambda_{i}-\tilde{\lambda}_{j}-2m\right)\right)+\mu\\
\rightarrow  0=&N\left(\frac{N_{f}}{2}-1\right)+2\sum_{j \in I_{+}}\coth\pi(\lambda_{i}-\lambda_{j})-N_{f}\sum_{j\in I_{-}}\tanh\pi\left(\lambda_{i}-\tilde{\lambda}_{j}\right)+\mu,
\end{align}
where we have taken the infinite mass limit in the last line. We also obtain $i \in I_{+}$ case by the same calculation. That is 
\begin{align}
0=N\left(1-\frac{N_{f}}{2}\right)+2\sum_{j\in I_{+}}\coth\pi(\tilde{\lambda}_{i}-\tilde{\lambda}_{j})-N_{f}\sum_{j \in I_{-}}\tanh\pi(\tilde{\lambda}_{i}-\lambda_{j})+\mu.
\end{align}
These two equations are same as the large $N$ saddle point equations of the following matrix model:
\begin{align}
\label{SYMadj3}
\int d^{N/2}\lambda d^{N/2}\tilde{\lambda}\delta\left(\sum_{i}\left(\lambda_{i}+\tilde{\lambda}_{i}\right)\right)\frac{e^{N(\frac{N_{f}}{2}-1)\sum_{i}\lambda_{i}}e^{N(1-\frac{N_{f}}{2})\sum_{i}\tilde{\lambda}_{j}}\prod_{i>j}\sinh^2\pi(\lambda_{i}-\lambda_{j})\sinh^2\pi(\tilde{\lambda}_{i}-\tilde{\lambda}_{j})}{\prod_{i,j}\cosh^{N_{f}}\pi(\lambda_{i}-\tilde{\lambda}_{j})}.
\end{align}
This matrix model is obtained from the S[U($\frac{N}{2}$)$\times$U($\frac{N}{2}$)] gauge theory with the $N_{f}$ massless bi-fundamental hypermultiplets and FI term deformation part related to the each U($\frac{N}{2}$), where we consider the indexes $i \in I_{+}$ corresponds to the indexes of one of the U($\frac{N}{2}$) and $i \in I_{-}$ corresponds to the indexes of the other U($\frac{N}{2}$). In fact this matrix model converges. Thus it is expected that the matrix model \eqref{SYMadj2} become \eqref{SYMadj3} in the infinite mass limit. 

When we consider $N_{f}=2$ case the FI term is vanishing. This may be the critical point which distinguishes the saddle point equation in the infinite mass limit with $N_{f}=2$ with other cases.  \footnote{
This FI term is related to the vector U(1) gauge group of U($\frac{N}{2}$)$\times$ U($\frac{N}{2}$). The axial U(1) gauge part is now forbidden by the condition $\sum_{i}\left(\lambda_{i}+\tilde{\lambda}_{i}\right)$. These FI terms can be interpreted as one-loop effects of massive adjoint fermions and massive gauginos of U($\frac{N}{2}$) \cite{CDFKS1,CDFKS2} }

Let us consider the following point of Coulomb branches of the original SU($N$) theory on the flat space:
\begin{align}
\label{Coulomb}
\sigma=\left(
\begin{array}{c|c}
-\frac{m}{2}\bf{1}_{\frac{N}{2}\times\frac{N}{2}}&\bf{0}\\\hline
\bf{0}&\frac{m}{2}\bf{1}_{\frac{N}{2}\times\frac{N}{2}}
\end{array}
\right).
\end{align} 
Then the theory is higgsed and the gauge group SU($N$) is broken to S[U($\frac{N}{2}$)$\times$U($\frac{N}{2}$)] and the $N_{f}$ massive adjoint hypermultiplets become effectively massless around this vacuum. This effective theory is same as the theory introduced above whose matrix model is given by \eqref{SYMadj3}. Then we can conclude that this split type solution corresponds to the point of the Coulomb branch \eqref{Coulomb}.

This result also means that we can not obtain the well-defined matrix model of pure SYM theory by regarding the massive hypermultiplets as the regularization. 

We have seen that the confinement solution is not a dominant saddle point solution for $k=0$. 
However, in the next subsection we propose the theory where the confinement solution is the dominant saddle point solution by adding the Chern-Simons term to the theory we considered here. With the help of CS term the eigenvalues feel the central force in the sense that we regard the saddle point equation as the E.O.M of the mechanics of the eigenvalues. Then the eigenvalues tend not to be split and the solution is forbidden. 

\subsection{A theory in which confinement solution is dominant}
Here we consider the SU($N$) Chern-Simons Yang-Mills theory\footnote{For U($N$) group, the similar discussions can be done.} with $N_{f}$ massive adjoint hypermultiplets. The matrix model is given by 
\begin{align}
Z=\frac{1}{N!}\int{d^{N}a}\delta(\sum_{i}^{N}a_{i})e^{i\pi k\sum_{i}a^2_{i}}\prod_{i>j}^{N}\frac{4\sinh^{2}\pi(a_{i}-a_{j})}{\left(2\cosh\pi(a_{i}-a_{j}+m)2\cosh\pi(a_{i}-a_{j}-m)\right)^{N_{f}}}.
\end{align}  
The saddle point equation is 
\begin{align}
0=2ika_{i}+2\sum_{j\neq i}\coth\pi(a_{i}-a_{j})-N_{f}\sum_{j}\left(\tanh\pi(a_{i}-a_{j}+m)+\tanh\pi(a_{i}-a_{j}-m)\right)+\mu,
\end{align}
where the first term of the R.H.S of the above equation is from the Chern-Simons term and causes the central force in the real and imaginary parts of the saddle point equation. With the help of the Chern-Simons term the split type solution no longer can exist. 

We will argue that the confinement solution is dominant 
for the large $N$ limit with $N \gg k $  with $\frac{k m}{N}$ finite and large.
To do this in detail we repeat the same argument in the previous subsection. The saddle point equation under the assumption that the solution which splits into two parts is given by
\begin{align}
\label{Chern-Simonssa1}
0=&2i\frac{k}{N}\left(\lambda_{i}-\frac{m}{2}\right)+\left(\frac{N_{f}}{2}-1\right)+\frac{2}{N}\sum_{j \in I_{-}}\coth\pi(\lambda_{i}-\lambda_{j})-\frac{N_{f}}{N}\sum_{j\in I_{+}}\tanh\pi\left(\lambda_{i}-\tilde{\lambda}_{j}\right)+\mu, \quad i\in I_{-},\\
\label{Chern-Simonssa2}
0=&2i\frac{k}{N} \left(\tilde{\lambda}_{i}+\frac{m}{2}\right)+\left(1-\frac{N_{f}}{2}\right)+\frac{2}{N}\sum_{j\in I_{+}}\coth\pi(\tilde{\lambda}_{i}-\tilde{\lambda}_{j})-\frac{N_{f}}{N}\sum_{j \in I_{-}}\tanh\pi(\tilde{\lambda}_{i}-\lambda_{j})+\mu,\quad i\in I_{+}.
\end{align}

Here, we consider a strong t' Hooft coupling limit $\frac{k}{N}\ll1$ and $\frac{m}{N}\ll1$ in order to make the confining solution valid.
If a combination of the parameters $\frac{km}{N}$ is small, then the Chern-Simons term can be neglected
and the split type solution is valid.
However, if $\frac{km}{N} = {\cal O}(N^0)$, the solution should be deformed and 
if $\frac{km}{N} $ is large we expect the split type solution can not exist.
We will not explicitly determine how large $\frac{km}{N} $ should be for the confinement phase
to be realized because it is difficult analytically. 
The numerical analysis shown below suggests that the critical value 
of $km/N $ is in $2 \lessapprox km/N \lessapprox 4$.
In figure 4, we show the density functions of some examples for these saddle point solutions.

\begin{figure}[h!]
 \begin{tabular}{cc}
 \begin{minipage}{0.5\hsize}
 \begin{center}
 \includegraphics[width=8cm]{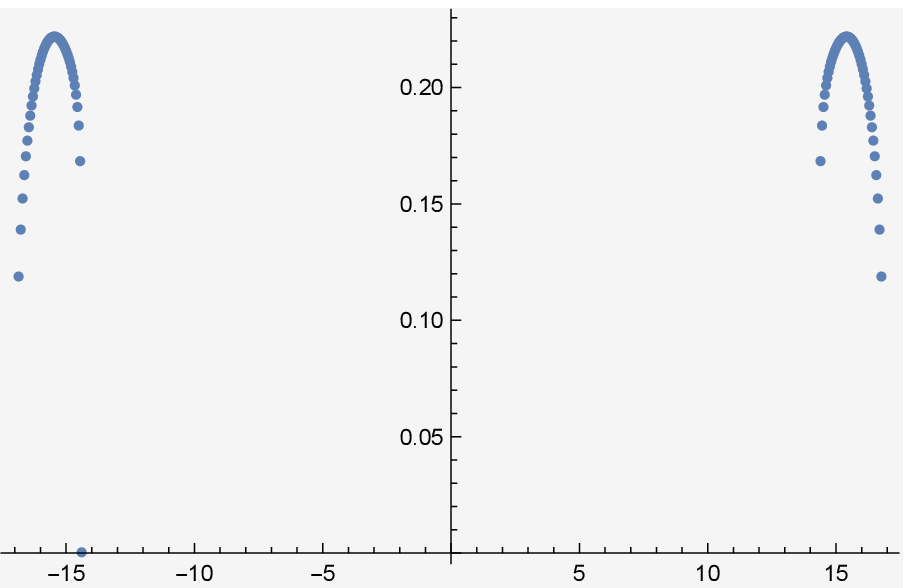}
 \label{nccomp3}
 \end{center}
 \end{minipage}
 \begin{minipage}{0.5\hsize}
 \begin{center}
\includegraphics[width=8cm]{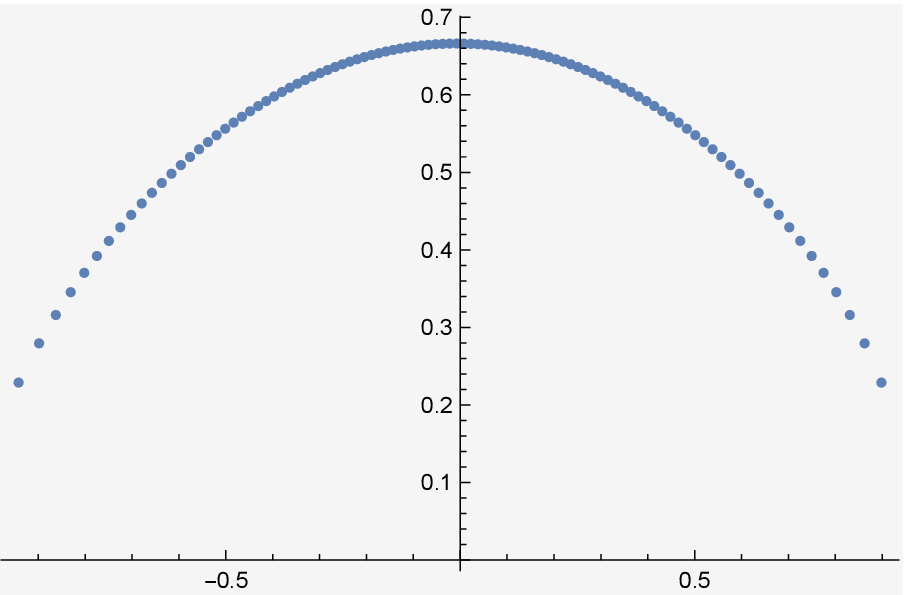}
 \label{nccomp4}
 \end{center}
 \end{minipage}
 \end{tabular}
\caption{These figure show that density functions of the real part of saddle point solution for $N=100$. The left one is with parameter $(k,m_a)=(10,10)$. The right one is with parameter $(k,m_a)=(150,10)$. The horizontal line means the value of the real part of the eigenvalue.}
\end{figure}

To show the behavior of the solution as the Chern-Simons level become large we summarize the value of the Wilson loop $W_{R}$ \eqref{SW} and the free energy $-\log|Z|$ from the numerical analysis in the following tables:

\begin{table}[h!]
  \begin{center}
    \begin{tabular}{|c|c|c|} \hline
       $(N,m)=(100.10)$ &  $W_{\text{R}}$ &  $-\log |Z|$ \\ \hline
      $k=10$   &$2.60356\times10^{15} - 2.87161\times10^{15} i$ &834190\\ 
      $k=20$  & $11.3867 - 0.10577 i$& 965491\\
      $k=30$  & $4.29052\times 10^{14} - 2.57048\times10^{15} i $ &53126\\
      $k=40$  & $22.56929 - 3.50511 i$ & 940792\\
      $k=50$  &$-0.914496 + 0.185339 i$ & 935889\\
      $k=60$  &$-3.46485 - 3.57006 i$ &939455\\
      $k=70$  & $1.0154 + 0.434397 i$ &935684 \\
      \hline
    \end{tabular} 
      \caption{The value of Wilson loop and free energy corresponding to the split solution discussed in the previous subsection from numerical analysis.}
  \end{center}
\end{table}

\begin{table}[h!]
\label{WLt2}
  \begin{center}
    \begin{tabular}{|c|c|c|} \hline
       $(N,m)=(100.10)$ &  $W_{\text{R}}$ &$-\log |Z| $\\ \hline
      $k=10$   &$-0.000341459 - 1.60786 i$&932650 \\ 
      $k=20$  & $-0.00133768 - 3.21552 i$&932824 \\
      $k=30$  & $-0.00289594 - 4.82279 i$&933113 \\
      $k=40$ &$-0.00484618 - 6.42953 i$ &933518 \\
      $k=50$&$0.0468883 - 7.95619 i$ &933976 \\
      $k=60$ &$ 0.0197115 - 9.61648 i$& 934642\\
      $k=70$ & $ -0.075207 - 8.80006 i$&934371\\
       \hline
    \end{tabular}
      \caption{The value of Wilson loop and free energy corresponding to the confinement type saddle solution \eqref{s2} from numerical analysis.}
  \end{center}
\end{table}
These values are evaluated by finding the solution of the saddle point equation numerically and using the saddle point approximation with the solution. From the result the value of the Wilson loop is drastically changed for $20 \lessapprox k \lessapprox 40$ when we take $N=100,m=10$. The solution of the saddle point equation no longer splits in this region
although we still call this the split type solution. The solution sits around the origin when $k$ is bigger than $30$, however the solution is different from the confinement solution.\footnote{
For $k=30$, the value of $-\log|Z|$ for the split solution seems to be strange.
We expect that there are some accidental reasons for this singular behavior 
because $N$ is finite.}
 When $k\geq 30$ there may be the solutions which do not depend on the mass, which also become close to the weak 't Hooft solution as the Chern-Simons level $k$ become large
\footnote{
Our claim is subtle because this argument is based on the numerical analysis of the saddle point equation. However, we can confirm that the behavior of the solution of the saddle point equation changes around $k=30$ and we could not find any counter example.
}.
 However, the value of the Wilson loop which is evaluated by the each of the two solutions is $\mathcal{O}(N^{0})$ 
which means there are some cancellations 
in  ${\rm Tr}_{R}$ in the definition of the Wilson loop
in the fundamental representation.
This is nothing but a characteristic property of the confinement phase.

In the tables, we also showed the $F=-\log|Z|$.
The values for the two different solutions 
are almost the same 
except $k=10$, for which the density function splits for one solution.  
Thus, we can not say which solution is dominant 
because the $1/N$ corrections will be important.
In order to decide which solution is dominant, we need 
to compute them numerically for larger $N$.
We hope to do it in near future.

Note that 
we have assumed that the probe approximation of the Wilson loop
is appropriate in this numerical computations.
Indeed, the values of $F$ are much larger than
the values of the logarithm of the Wilson loop.
This fact justifies the probe approximation.


\vspace{1cm}




\newpage 
\begin{thebibliography}{999}
\parskip=-2pt


\bibitem{Pestun}
  V.~Pestun,
  ``Localization of gauge theory on a four-sphere and supersymmetric Wilson loops,''
  Commun.\ Math.\ Phys.\  {\bf 313} (2012) 71
  doi:10.1007/s00220-012-1485-0
  [arXiv:0712.2824 [hep-th]].

\bibitem{Witten}
  E.~Witten,
  ``Topological Quantum Field Theory,''
  Commun.\ Math.\ Phys.\  {\bf 117} (1988) 353.
  doi:10.1007/BF01223371

\bibitem{Nekrasov}
  N.~A.~Nekrasov,
  ``Seiberg-Witten prepotential from instanton counting,''
  Adv.\ Theor.\ Math.\ Phys.\  {\bf 7} (2003) no.5,  831
  doi:10.4310/ATMP.2003.v7.n5.a4
  [hep-th/0206161].




\bibitem{KWY1} 
  A.~Kapustin, B.~Willett and I.~Yaakov,
  ``Exact Results for Wilson Loops in Superconformal Chern-Simons Theories with Matter,''
  JHEP {\bf 1003}, 089 (2010)
  [arXiv:0909.4559 [hep-th]].


\bibitem{Jafferis}
  D.~L.~Jafferis,
  ``The Exact Superconformal R-Symmetry Extremizes Z,''
  JHEP {\bf 1205} (2012) 159
  doi:10.1007/JHEP05(2012)159
  [arXiv:1012.3210 [hep-th]].


\bibitem{Hama}
  N.~Hama, K.~Hosomichi and S.~Lee,
  ``Notes on SUSY Gauge Theories on Three-Sphere,''
  JHEP {\bf 1103} (2011) 127
  doi:10.1007/JHEP03(2011)127
  [arXiv:1012.3512 [hep-th]].

\bibitem{IS} 
  K.~Intriligator and N.~Seiberg,
  ``Aspects of 3d N=2 Chern-Simons-Matter Theories,''
  JHEP {\bf 1307}, 079 (2013)
  [arXiv:1305.1633 [hep-th]].

\bibitem{Witten1} 
  E.~Witten,
  ``Supersymmetric index of three-dimensional gauge theory,''
  In *Shifman, M.A. (ed.): The many faces of the superworld* 156-184
  [hep-th/9903005].
  
 \bibitem{Ohta}
  K.~Ohta,
  ``Supersymmetric index and s rule for type IIB branes,''
  JHEP {\bf 9910} (1999) 006
  doi:10.1088/1126-6708/1999/10/006
  [hep-th/9908120].
  
 \bibitem{NST2} 
  T.~Nosaka, K.~Shimizu and S.~Terashima,
  ``Mass Deformed ABJM Theory on Three Sphere in Large N limit,''
  JHEP {\bf 1703}, 121 (2017)
  [arXiv:1608.02654 [hep-th]].


\bibitem{HNST} 
  M.~Honda, T.~Nosaka, K.~Shimizu and S.~Terashima,
  ``Supersymmetry Breaking in a Large $N$ Gauge Theory with Gravity Dual,''
  arXiv:1807.08874 [hep-th].

\bibitem{NST1} 
  T.~Nosaka, K.~Shimizu and S.~Terashima,
  ``Large N behavior of mass deformed ABJM theory,''
  JHEP {\bf 1603}, 063 (2016)
  doi:10.1007/JHEP03(2016)063
  [arXiv:1512.00249 [hep-th]].

\bibitem{Hosomichi} 
  N.~Hama, K.~Hosomichi and S.~Lee,
  ``SUSY Gauge Theories on Squashed Three-Spheres,''
  JHEP {\bf 1105}, 014 (2011)
  [arXiv:1102.4716 [hep-th]].

\bibitem{Kapustin}
  A.~Kapustin, B.~Willett and I.~Yaakov,
  ``Nonperturbative Tests of Three-Dimensional Dualities,''
  JHEP {\bf 1010} (2010) 013
  doi:10.1007/JHEP10(2010)013
  [arXiv:1003.5694 [hep-th]].


\bibitem{3d}
  K.~Hosomichi,
  ``A Review on SUSY Gauge Theories on $\mathbf{S^3}$,''
  doi:10.1007/978-3-319-18769-3 10
  arXiv:1412.7128 [hep-th].

\bibitem{BT} 
  A.~G.~Bytsko and J.~Teschner,
  ``Quantization of models with non-compact quantum group symmetry: Modular XXZ magnet and lattice sinh-Gordon model,''
  J.\ Phys.\ A {\bf 39}, 12927 (2006)
  doi:10.1088/0305-4470/39/41/S11
  [hep-th/0602093].
  
   
\bibitem{Hatsuda} 
  Y.~Hatsuda,
  ``ABJM on ellipsoid and topological strings,''
  JHEP {\bf 1607}, 026 (2016)
  doi:10.1007/JHEP07(2016)026
  [arXiv:1601.02728 [hep-th]].
 
\bibitem{CKW1} 
  C.~Closset, H.~Kim and B.~Willett,
  ``Supersymmetric partition functions and the three-dimensional A-twist,''
  JHEP {\bf 1703}, 074 (2017)
  doi:10.1007/JHEP03(2017)074
  [arXiv:1701.03171 [hep-th]].
 
 
\bibitem{CKW2} 
  C.~Closset, H.~Kim and B.~Willett,
  ``Seifert fibering operators in 3d $\mathcal{N}=2$ theories,''
  arXiv:1807.02328 [hep-th].


\bibitem{AHW} 
  I.~Affleck, J.~A.~Harvey and E.~Witten,
  ``Instantons and (Super)Symmetry Breaking in (2+1)-Dimensions,''
  Nucl.\ Phys.\ B {\bf 206}, 413 (1982).
  doi:10.1016/0550-3213(82)90277-2
  
\bibitem{MN} 
  T.~Morita and V.~Niarchos,
  ``F-theorem, duality and SUSY breaking in one-adjoint Chern-Simons-Matter theories,''
  Nucl.\ Phys.\ B {\bf 858}, 84 (2012)
  [arXiv:1108.4963 [hep-th]].
  
   
\bibitem{Suyama1} 
  T.~Suyama,
  ``Supersymmetry Breaking and Planar Free Energy in Chern-Simons-matter Theories,''
  arXiv:1405.7469 [hep-th].
 
  
    
\bibitem{HLLLP} 
  K.~Hosomichi, K.~M.~Lee, S.~Lee, S.~Lee and J.~Park,
  ``N=5,6 Superconformal Chern-Simons Theories and M2-branes on Orbifolds,''
  JHEP {\bf 0809}, 002 (2008)
  doi:10.1088/1126-6708/2008/09/002
  [arXiv:0806.4977 [hep-th]].
\bibitem{GRVV} 
  J.~Gomis, D.~Rodriguez-Gomez, M.~Van Raamsdonk and H.~Verlinde,
  ``A Massive Study of M2-brane Proposals,''
  JHEP {\bf 0809}, 113 (2008)
  doi:10.1088/1126-6708/2008/09/113
  [arXiv:0807.1074 [hep-th]].
    

\bibitem{M} 
  R.~C.~Myers,
  ``Dielectric branes,''
  JHEP {\bf 9912}, 022 (1999)
  doi:10.1088/1126-6708/1999/12/022
  [hep-th/9910053].


\bibitem{ID} 
  Y.~Imamura and D.~Yokoyama,
  ``N=2 supersymmetric theories on squashed three-sphere,''
  Phys.\ Rev.\ D {\bf 85}, 025015 (2012)
  doi:10.1103/PhysRevD.85.025015
  [arXiv:1109.4734 [hep-th]].

\bibitem{TW} 
  C.~Toldo and B.~Willett,
  ``Partition functions on 3d circle bundles and their gravity duals,''
  JHEP {\bf 1805}, 116 (2018)
  doi:10.1007/JHEP05(2018)116
  [arXiv:1712.08861 [hep-th]].

\bibitem{GK} 
  D.~Gang and N.~Kim,
  ``Large $N$ twisted partition functions in 3d-3d correspondence and Holography,''
  arXiv:1808.02797 [hep-th].


\bibitem{GW} 
  D.~Gaiotto and E.~Witten,
  ``Supersymmetric Boundary Conditions in N=4 Super Yang-Mills Theory,''
  J.\ Statist.\ Phys.\  {\bf 135}, 789 (2009)
  [arXiv:0804.2902 [hep-th]].

\bibitem{CDFKS1} 
  C.~Closset, T.~T.~Dumitrescu, G.~Festuccia, Z.~Komargodski and N.~Seiberg,
  ``Contact Terms, Unitarity, and F-Maximization in Three-Dimensional Superconformal Theories,''
  JHEP {\bf 1210}, 053 (2012)
  [arXiv:1205.4142 [hep-th]].
\bibitem{CDFKS2} 
  C.~Closset, T.~T.~Dumitrescu, G.~Festuccia, Z.~Komargodski and N.~Seiberg,
  ``Comments on Chern-Simons Contact Terms in Three Dimensions,''
  JHEP {\bf 1209}, 091 (2012)
  doi:10.1007/JHEP09(2012)091
  [arXiv:1206.5218 [hep-th]].


  
  





\end{thebibliography}
\end{document}